\documentclass[%
 reprint,
 amsmath,amssymb,
 aps,
]{revtex4-2}

\usepackage{graphicx}
\usepackage{dcolumn}
\usepackage{bm}

\usepackage[T1]{fontenc} 
\usepackage{booktabs} 
\usepackage{ dsfont }
\usepackage{float}

\usepackage{calrsfs}
\DeclareMathAlphabet{\pazocal}{OMS}{zplm}{m}{n}
\usepackage{xcolor}
\usepackage{pict2e}
\usepackage[percent]{overpic}
\usepackage{ upgreek }
\usepackage{xfrac}
\usepackage{abraces}
\usepackage{braket}
\usepackage{stmaryrd}
\usepackage{amsmath,empheq}
\usepackage{nicefrac}

\begin{document} 

\preprint{APS/123-QED}

\title{Axion-photon multimessenger astronomy with giant flares}
\author{Javier De Miguel}
 \email{javier.miguelhernandez@riken.jp}
\author{Chiko Otani}%
 \affiliation{
The Institute of Physical and Chemical Research (RIKEN), \\
Center for Advanced Photonics, 519-1399 Aramaki-Aoba, Aoba-ku, Sendai, Miyagi 980-0845, Japan}


\date{\today}

\begin{abstract}
We treat prospects for multimessenger astronomy with giant flares (GFs), a rare transient event featured by magnetars that can be as luminous as a hundred of the brightest supernovae ever observed. The beamed photons could correlate with an axion counterpart via resonant conversion in the magnetosphere. In a  realistic parameter space, we find that the sensitivity limit to galactic GFs for currently viable experiments is $\mathrm{g}_{\phi \gamma}\!\gtrsim\!\mathrm{several}\!\times\!10^{-13}$ GeV$^{-1}$ \& $\mathrm{g}_{\phi e}\!\gtrsim\!\mathrm{few}\!\times\!10^{-12}$. We rule out the compatibility of axion flares with the recent XENON1T excess only due to the time persistence of the signal.
\end{abstract}

\maketitle


\section{Background}
The Quantum Chromodynamics (QCD) axion \cite{PhysRevLett.40.223, PhysRevLett.40.279} is a hypothetical pseudo-scalar boson that arises from the dynamic solution to the charge and parity (CP) problem in the strong interaction \cite{PhysRevLett.38.1440}. The axion mass — $\mathrm{m}_{\phi}$ — and coupling constants — $\mathrm{g}$ — are unknown in spite of the remarkable effort realized to find axion in the parameter space at which that can simultaneously solve the CP problem and the dark matter (DM) enigma \cite{ABBOTT1983133, DINE1983137, PRESKILL1983127,PhysRevD.98.030001}. Relevant interactions of axion-like particles to Standard Model (SM) particles for this work are

\begin{eqnarray}
\pazocal{L}^{\mathrm{int}} =\label{Eq.1} -\frac{1}{4}\mathrm{g}_{\phi\gamma} F_{\!\mu \nu} \tilde{F}^{\mu \nu}\!\phi + i \mathrm{g}_{\phi e} \overline{\psi}\,\gamma^{_5} \psi \phi 
\,,
\end{eqnarray}
where the first term on the right-hand side represents the effective axion-photon Lagrangian density and the second term the axion-electron coupling; being $\phi$ the axion field, $F^{\mu \nu}$ the photon field, $\psi$ the electron field and $\gamma_{\!_5}$ the usual matrix. 

\section{Axion-photon mixing in astrophysical environments}
\subsection{General dispersion relations}
Aided by axion-electrodynamics \cite{Wilczek:1987mv} for the linearization of the equations of motion from Eq.$\,$\ref{Eq.1}, a set of Schrödinger-like dispersion relations emerges that can be solved analytically in idealized astrophysical boundaries  \cite{PhysRevD.37.1237}

\begin{equation}
\left[\omega \!+\!\!
\begin{pmatrix} \Delta_{\!\perp}\!\!+\!\Delta_{Q_{\parallel}} & \Delta_{\!R} & \Delta_{\mathrm{M}_{x}}\\ \Delta_{\!R} & \Delta_p\!+\!\Delta_{_{\parallel}} \!\!+\!\Delta_{Q_{\!\perp}} & \Delta_{\mathrm{M}_{y}} \\ \Delta_{\mathrm{M}_{x}} & \Delta_{\mathrm{M}_{y}} & \Delta_{\phi} \end{pmatrix}\!\!-\!i 
\partial z \right] \!\!\!\begin{pmatrix} \,\mathrm{E}_{\!\perp} \\  \mathrm{E}_{_{\parallel}} \\ \phi \!\!\end{pmatrix} \!\!= \!0 \,.
\label{Eq.4}
\end{equation}
The interaction term $\pazocal{L}^{\phi\gamma}\!\!=\!\mathrm{g}_{\phi\gamma}\mathrm{E}\!\cdot\! \mathrm{B}$ is obtained from Eq.$\,$\ref{Eq.1}
being E the photon and B the external magnetic field, establishing a preferential polarization. Now, obviating the Faraday rotation term — $\Delta_{\!R}$ — the perpendicular component $\mathrm{E}_{\!\perp}$ decouples and the mixing is reduced to a 2D system given by the low-right sector in the second term of the left-side hand of Eq.$\,$\ref{Eq.4}. The term $\Delta_{_{\parallel}}\!\!=\!\frac{1}{2}\omega (\mathrm{n_{_{\parallel}}}\!\!-\!1)$ vanishes in the weak dispersion limit, since the refraction index is $\mathrm{n_{_{\parallel}}}\!\!\!\sim\!\!1$ for relativistic axions. The quantum-electrodynamics (QED) vacuum refraction parameter — $\Delta_{Q_{}}$— can be treated independently. We have defined the refraction parameter $\Delta_{p}\!\!=\!\!-\nicefrac{\omega^2_{\!p}}{2\omega}$, with $\omega_p\!=\! (e^2 n_{e}/m_e\varepsilon_{_0})^{\!\nicefrac{1}{2}}$ the characteristic frequency of the plasma, and the axion mass parameter $\Delta_{\phi}\!\!=\!\!-\nicefrac{\mathrm{m}^2_{\!\phi}}{2\omega}$. The coupling term is $\Delta_{\mathrm{M}}\!\!=  \!\mathrm{g}_{\phi\gamma}\mathrm{B}/2$. We also define $\Delta \mathrm{k}^2\!=\!
(\Delta_p+\Delta_{Q}- \Delta_{\phi})^2+4\Delta^2_{\mathrm{M}}$ and the mixing angle that holds the diagonalization, $ \mathrm{tan}(2\theta_m\!)\!=\!2\Delta_{\mathrm{M}}/(\Delta_p-\Delta_{\phi})$.

\subsection{The role of the plasma density in the relative weight of QED corrections}
It has been pointed out that QED vacuum polarization effect acts as a suppressor of the axion-(keV)photon cross-section in a minimal model of neutron star adopting the Goldreich-Julian (GJ) corotation density — $n_{_{\mathrm{GJ}}}\!=\!4\pi\varepsilon_{\!_0} \!B(r)(eP)^{-1}$ — \cite{PhysRevD.37.1237, 1969ApJ...157..869G}. However, a pair multiplicity factor, defined $\kappa\!=\!n_{e}/n_{_\mathrm{{GJ}}}$, emerges in modern approaches beyond the GJ paradigm \cite{demiguel2021superdense}. The ratio of plasma effects to QED vacuum effects, manifested through the Euler-Heisenberg term $\pazocal{L}_{\mathrm{QED}}\!=\!\frac{\alpha^{2}}{90m^{4}_{\!e}}[\left(F_{\!\mu\nu}F^{\mu\nu}\right)^{2}\!+\frac{7}{4}(F_{\!\mu\nu}\tilde{F}^{\mu\nu})^{2}]$, scales with the density profile across the \textit{gas} \cite{PhysRevD.37.1237, PhysRevD.97.123001}

\begin{equation}
\frac{\Delta_{p}}{\Delta_{Q_{\!\perp}}} = 5\!\times\!10^{-10}\left(\!\frac{\mathrm{ keV}}{\omega}\!\right)^{\!\!2}\!\frac{\mathrm{10^{8} [T]}}{ B(r)}\frac{\mathrm{1[s]}}{P}\,\kappa \, .
\label{Eq.0}
\end{equation}

From Eq.$\,$\ref{Eq.0}, it follows that the tension with QED results progressively relaxed above $\kappa\!\!\gtrsim\!\!10^{5}$ at relatively large distances from the star for keV photons, while $\Delta_{Q}$ gradually vanishes near its surface for $\kappa\!\gtrsim\!\!10^{8}$. In the case of very energetic transitory events featured by magnetars, the charge density near the star surface is $n_{e}\!\!\sim\!\!10^{30}$cm$^{-3}$ \cite{Roberts:2021udn}, many orders over the occupation number predicted by the GJ model. Although the density profile through the \textit{atmosphere} of the `fireball' is barely known, simulations suggest that pair multiplicity factors $\kappa\!\!\lesssim\!\!10^{10}$ are tenable at an altitude of few stellar radii \cite{2015ApJ...815...45Y}. That renders resonant conversion efficient in the parameter space dominated by plasma effects; including highly magnetic neutron stars, where $10^{4-5}\!\!\lesssim\!\kappa\!\lesssim\!\!10^{7}$ \cite{Guepin:2019fjb, demiguel2021superdense}, against previous expectations. 

\subsection{Magnetar axion}
Magnetars are compact stellar remnants endowed with extreme magnetic fields, typically about $10^{10-11}$ T, a `canonical' mass around 10-20 solar masses and a rotation period of the order of seconds \cite{Duncan1992FormationOV}. In the standard picture, their persistent emission mechanisms involve pair production and magnetic acceleration, catalyzing the migration of charges to higher altitudes from the star to give raise to curvature radiation, synchrotron emission and inverse Compton scattering \cite{1975ApJ...196...51R, 1979ApJ...231..854A}. The beamed photons could be converted to axion through non-adiabatic resonant mechanism at their path across the non-relativistic, `cold', plasma of the magnetosphere  \cite{PhysRevD.74.123003, Perna:2012wn, PhysRevD.97.123001, PhysRevD.102.023504, Zhuravlev:2021fvm, demiguel2021superdense}. In the aligned-rotator approximation, where the magnetic axis and the rotation axis — $\hat{z}$ — are superimposed, the axion-photon oscillation amplitude is derived from introducing the resonance condition $\Delta_{p}-\Delta_{\phi}\!=\!0$ in Eq.$\,$\ref{Eq.4}, yielding a conversion probability $\mathord{\mathrm{P}}_{\!\!\!\gamma \!\!\;\phi}\!=\Delta\!^2\mathrm{k}/(2\frac{\partial}{\partial z}\Delta_p$). Since outgoing photons present a gradient $\frac{\partial}{\partial z}\Delta_p\! \sim\! \mathrm{m^2_{\phi}}\omega^{-1}z_{c}^{-1}$, it is straightforward to obtain \cite{demiguel2021superdense}

\begin{equation}
\mathord{\mathrm{P}}_{\!\!\!\gamma \!\!\;\phi}= \frac{\mathrm{g}^2_{\phi\gamma} B^2(z_{c}) \,\omega  \,z_c}{\mathrm{m^2_{\phi}}} \,.
\label{Eq.5}
\end{equation}

In Eq.$\,$\ref{Eq.5}, $B(r)\!=\!B_0(\mathrm{R}/r)^3$ expresses the magnetic field in the dipole approximation, stationary as a result from an aligned rotator, as a function of the distance from the surface of the star, $z_c$ is the conversion altitude, $\omega$ the pulse of the photon.

\begin{figure*}[t] \centering
\includegraphics[width=.90\textwidth]{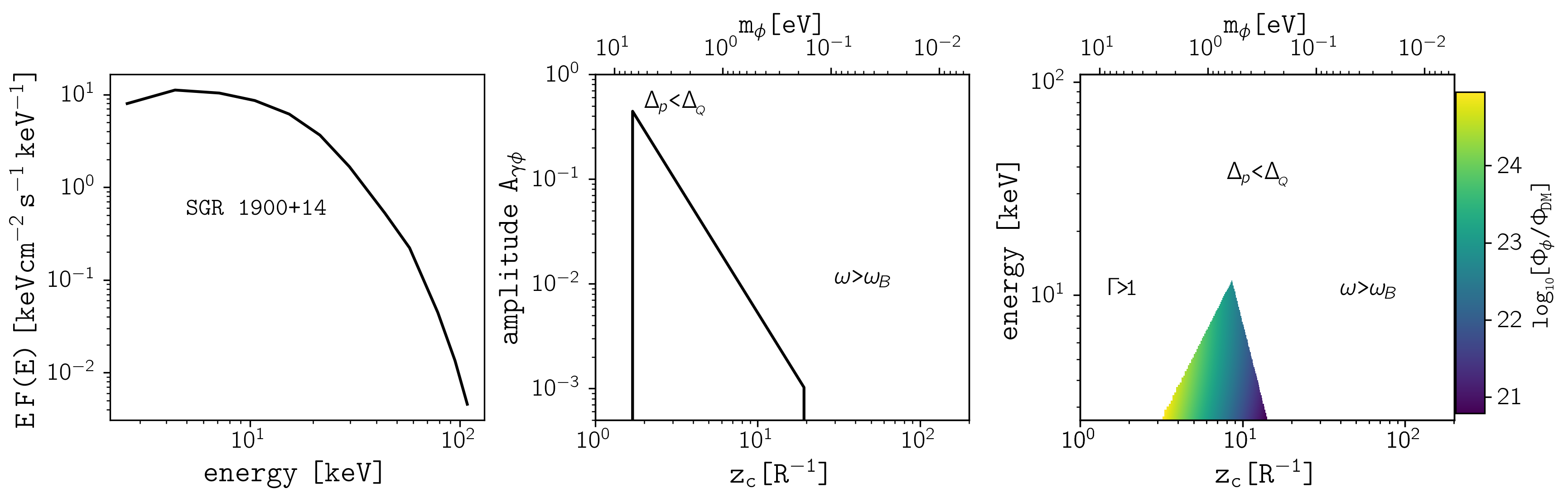}
\caption{\label{fig:epsart}Left: Deconvolved spectra of an intermediate flare (IF) from SGR 1900+14 occurred on 2001 July 2 with a duration of several seconds. It combines FREGATE data between 7–150 keV and WXM data between 2–25 keV. The bolometric luminosity was $\gtrsim\!\!10^{41}$ $\!$erg $\!$s$^{-1}$. We consider and therefore represent a 50\% fraction of the incoming beam polarized in the O-mode. Adapted from \cite{2004ApJ...616.1148O}. Middle: Photon-to-axion oscillation amplitude at $\!\sim$keV energy as a function of the resonance altitude — $\mathrm{z_c}$ — in units of the stellar radius or axion mass, $\mathrm{m_{\phi}}$. Right: Axion flux density — $\Phi_{\!\phi}$ — triggered by the IF in units of ambient dark matter (DM) density — $\Phi_{\!\mathrm{_{DM}}}$ — over altitude or axion mass. The pair multiplicity factor over the Goldreich-Julian (GJ) corotation density is defined $\kappa\!=\!n_{e}/n_{_\mathrm{{GJ}}}$ and set $\kappa\!\sim\!\!10^8$. The rotation period $P\!\!\sim\!\!1$ $\!$s, $B_{\!_0}\!\!\sim\!\!10^{10}$ T and $\mathrm{R}\!\sim\!10^6$ cm are canonical dimensions. We adopt QCD axion coupling strength. The non-adiabatic resonant model is ruled out in sectors where O-modes become evanescent, or equivalently $\omega_p\!>\!\omega$; $\Upgamma \!\gtrsim\!1$ regions, being $\Upgamma$ the adiabaticity factor of the system; `weak-field' zones, with $\omega\!\gtrsim\!\omega_{\!_B}$, being $\omega_{\!_B}\!\!=\!e B(z_c)/m_e$ the gyrofrequency; sectors where QED vacuum polarization effect governs the dispersion relations, or $\Delta_p\!<\!\Delta{_Q}$. In a consequence, axions with $\mathrm{m_{\phi}}\!\gtrsim\!10^{-1}$ eV mix efficiently with 1-10 keV photons.}
\label{Fig.1}
\end{figure*}

\section{Magnetar flares and their axion counterpart}
\subsection{Magnetar flares}
The so-called anomalous X-ray pulsars (AXPs) and soft gamma-ray repeaters (SGRs) today are thought to be transient events from magnetars. Their strong outbursts are frequently referred short-bursts (SBs), intermediate flares (IFs) and giant flares (GFs), whose more general characteristics are summarized in Table \ref{table1}. 
\begin{table}[b]
\caption{ Properties of X/$\gamma$-ray emission from magnetars.}
\begin{ruledtabular}
\begin{tabular}{ccccc}
\label{table1}
\textrm{Feature}&
\textrm{Object}&
\multicolumn{1}{c}{\textrm{Time-scale\footnote{Duration of the peak.}[s]}}&
\textrm{L\footnote{Bolometric luminosity; isotropic flux.} $\!$[erg$\!$ s$^{-1}$}]\\
\colrule \rule{0pt}{2.5ex}
Quiescent & Magnetar & - & $10^{33}$\footnote{Typical value with a variability of several orders of magnitude.} \\
SB & AXP\&SGR & 0.2--1 & $10^{39-41}$\\
IF & AXP\&SGR & 1--40 & $10^{41-43}$\\
GF & SGR & <0.5+tail\footnote{A few-hundred seconds scattering tail with a lower luminosity follows the main peak.} & $10^{44-47}$\\
\end{tabular}
\end{ruledtabular}
\end{table}
If the quiescent emission mechanisms of magnetars are poorly understood, the scenario for transient events is even more complex. Rapid magnetic field reconfiguration is thought to be an important part of the outbursts. The released magnetic energy is converted then to thermal energy in the magnetosphere — cf. \cite{Boggs:2006uk} for solved non-thermal spectral features —. However, the triggering mechanisms, and the exact role of the magnetic field remain unresolved. The time-scale of the magnetic reconnection — $t_{\mathrm{rec}}\!\sim\!L/V$, where $L$ is the height of the reconnection point and $V$ the inflow velocity — should be much larger than the typical time-scale of the system, $\tau_{\!s}\!\sim \mid\!\! \omega-\mathrm{m}_{\phi}\!\!\mid^{-1}$, to render plausible the conversion mechanism described by Eq.$\,$\ref{Eq.5}. Within this scenario, reasonable for a relatively high frequency or a sufficiently high distance from the star, we have $\tau_{\!s}\!\!<\!<\!t_{\mathrm{rec}}$ and the oscillation takes place in a pseudo-stationary regime. We include the spectra of an IF with energy-scale about $10^{41}$erg $\!$s$^{-1}$ in Fig.$\,$\ref{Fig.1} left. The correlated axion flare is quantified in units of the typical occupation of virialized dark matter in the Galactic halo, around $10^{13}$ $\!$keV$\,\!$cm$^{-2}$\,\!s$^{-1}$.

We will confine our attention throughout the following sections to the study of axion giant flares (AGFs) triggered by GFs. Three `galactic' GFs have been observed: from SGR 0526-66, actually hosted by Large Magellanic Cloud (LMC), in 1979 \cite{Mazets1979ObservationsOA}; from SGR 1900+14 in 1998 \cite{Hurley:1998ks} and, the most luminous to date, from SGR 1806-20 in 2004 \cite{2005Natur.434.1098H, 2005Natur.434.1107P}. Although there is consensus that GFs are rare events, the exact rate is source of controversy and is susceptible to observational bias \cite{HURLEY20111337}. Statistics in \cite{Burns_2021} suggests that the rate for events above $4\!\times\!10^{44}$ $\!$erg is $\sim\!\!3.8\!\times\!10^5$ Gpc$^{-3}$ $\!$yr$^{-1}$, resulting in a event rate per magnetar of the order of $\lesssim\!\!$ 0.02 yr$^{-1}$. These estimates rely on the strong evidence that 1-20\% of the gamma-ray burst (GRB) sample are indeed extragalactic GFs, whose tails are undetectable due to the large distance \cite{Roberts:2021udn}. 

In that manner, the catalogue of confirmed GFs is poor and, in addition, their spectral features are not adequately characterized due to saturation of the detectors and other systematics. Therefore, we fit and extend to low frequency a black-body (BB) curve at a temperature $k_{\!_B}\! T_{_{\mathrm{\!BB}}}\!\!\!\sim$175 $\!$keV, with a fluence of 1.36 erg $\!$cm$^{-2}$ of $>$30 keV photons on Earth consistent with the SGR 1806-20 event \cite{2005Natur.434.1098H}, to be employed in simulations throughout the following sections. Notwithstanding that empirical data are not available when referring to the polarization of the GF, it is predicted that the strong magnetic field in magnetars linearly polarizes the photons in two normal modes, i.e., ordinary (O) — with the E field lying on the k-B plane — and extraordinary (X). The weight of the O-mode, which efficiently mix with axion, was established in a conservative 0.5 fraction for our estimates after consulting the literature \cite{2002MNRAS.332..199L, Yang:2015zhz, vanPutten:2016qri, Taverna:2017ftz}.

\subsection{The flight of axion to Earth}
Obviously, both the photon and axion flux suffer from the inverse-square law with distance. Moreover, axions with a rest mass $\mathrm{m_{\phi}}\!\!\lesssim\!\!10^{-5}$ $\!$eV  crossing through several cosmic magnetic domains, with a typical length scale — $s$ — of the order of Mpc, are reconverted efficiently into photons; first across their outgoing path through the intergalactic medium (IGM), and then upon reaching the Milky Way (MW) \cite{PhysRevLett.76.2633, PhysRevD.77.063001, PhysRevD.84.125019}. The photon-to-axion conversion probability, derived from Eq.$\,$\ref{Eq.4}, is read $\mathord{\mathrm{P}}_{\!\!\!\gamma \!\phi}\!=\!4\Delta^2_{\mathrm{M}} \mathrm{sin}^2(\ell\Delta\mathrm{k}/2)/\Delta\!^2\mathrm{k}$. In the strong-coupling limit, the inequality $(\Delta_p\!+\!\Delta_Q\!-\!\Delta_{\phi})^2\!<\!<\!4\Delta^2_{\mathrm{M}}$ is obtained, resulting in the well-known expression \cite{PhysRevD.39.2089}

\begin{equation}
\mathord{\mathrm{P}}_{\!\!\phi \!\!\;\gamma}= \frac{\mathrm{g}^2_{\phi\gamma} \,B^2 \,\ell^2}{4}\,.
\label{Eq.6}
\end{equation}

From a panoramic perspective, our galaxy takes the form of a thin disk of $\sim$30 kpc length  — i.e., $\ell\!\!<\!<\!\!s$ — with a magnetic field of the order of a fraction of nT, and a low electron density, about $10^{-3}$ $\!$cm$^{-3}$. Crucially, from Eq.$\,$\ref{Eq.6} it follows that only a marginal fraction of the magnetar axions released in the vicinity of the MW are reconverted into photons during their approximation to Earth. In other words, the astroparticle flux density at origin, determined by Eq.$\,$\ref{Eq.5}, approximately persists.

Photons are massless through the IGM. Axion is massive. In contrast to the plasma that envelopes the magnetosphere, where the resonant conversion takes place at $\omega_{\!p}\!\!\sim\!\! \mathrm{m}_{\phi}$, the relation $\hbar\omega\!\!\sim\!\!\upgamma\mathrm{m}_{\phi}$ is maintained through the vacuum, being $\upgamma$ the Lorentz factor. The axion velocity is $\beta_{\!\phi}\!=\!\sqrt{1-1/\upgamma^2}$. In that form, the delay time, at keV energies, for values of the Lorentz factor between $10^{1-7}$ and pair multiplicity factor $10^{2}\!\!\lesssim\!\kappa\!\lesssim\!10^{9}$ varies between about a year per traveled kpc and a few milliseconds. That would render photon-axion multimessenger astronomy constrained to a human-scale time frame and would motivate the tracking of confirmed outbursts for relatively massive axions.

\begin{figure}[t]\centering
\includegraphics[width=.38\textwidth]{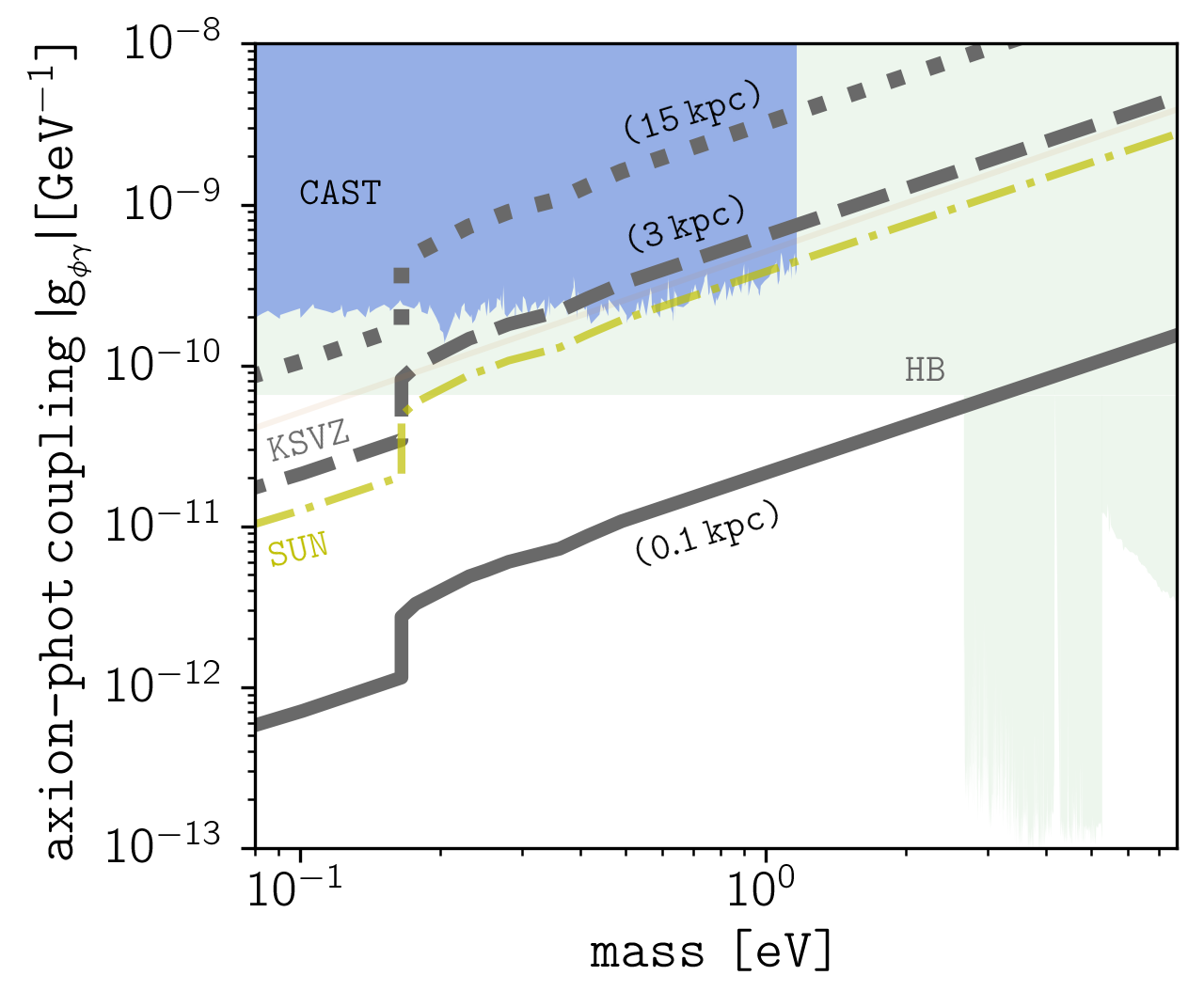}
\caption{Limit on the sensitivity to axion flares by realistic helioscope-like detectors, based on the inverse Primakoff-effect. The CAST helioscope limit is established at 95\% confidence level (CL) \cite{CAST:2017uph}. The horizontal branch (HB) exclusion region is established by indirect methods at 95\% CL \cite{Ayala:2014pea}. Astro-bounds established by indirect methods are also included in green \cite{Regis:2020fhw, Grin:2006aw}. The light brown solid line represents the KSVZ axion. The dot-dash yellow line projects new-generation helioscope sensitivity at 95\% CL from a campaign lasting several years \cite{Irastorza:2011gs}, while the duration of the giant flare is set $t\!=\!0.125\!$ s. Dark lines correspond to direct detection of GFs with 15 — dotted —, 3 — dashed — and 0.1 — solid — kpc distance, at 95\% CL. The comparison is performed by integration of the spectral functions between 1-10 keV, where the detectors are more sensitive coinciding with the maximal flux of solar axion.}
\label{Fig.4}
\end{figure}

\section{Observable effects}

\subsection{Detection of axion flares through inverse Primakoff-effect}
The solar axion spectra is formed by interactions of axion with SM particles in the internal plasma of the star. A spectra dominated by Primakoff-effect presents a maximal flux about $10^{9}\!$ $\!$cm$^{-2}\!$ $\!$s$^{-1}$ at 3-4 keV for an axion-photon coupling strength $\mathrm{g}_{\phi\gamma}\!\!\sim\!\! 10^{-11}$ GeV$^{-1}$. A solar spectra shaped by atomic recombination and deexcitation, bremsstrahlung, and Compton (ABC) processes peaks a similar fluence around 1 keV for an axion-electron coupling constant $\mathrm{g}_{\phi\mathrm{e}}\!\!\sim\!\! 10^{-13}$. Differently from magnetar axion, where one finds from Eq.$\,$\ref{Eq.5} that the QCD axion flux is enhanced for X-ray photons, the emitted solar flux scales with $\mathrm{g}^{2}$ mitigating the emission of axion-like particles (ALPs) with a low coupling strength, including both KSVZ  \cite{PhysRevLett.43.103, Shifman1980CanCE} and DFSZ \cite{DINE1981199, osti_7063072} axion.

Helioscopes are directional detectors tracking the Sun \cite{PhysRevLett.51.1415, PhysRevLett.69.2333, Moriyama:1998kd, PhysRevLett.94.121301}. Their working-principle relies on inverse Primakoff axion-to-photon conversion — $\phi+\gamma_{_{\!\mathrm{virt}}}\!\!\!\rightarrow\!\gamma$ — in a magnetized cryostat equipped with photomultiplier tubes (PMTs). The cross-section of helioscopes in the limit $q\ell\!<\!<\!1$ is given by Eq.$\,$\ref{Eq.6}, being $q$ the transfer of momenta between the axion and the photon and $\ell$ the length of flight across the vessel. The product of the inductance of the magnetic field and the length/area is restricted in a physical implementation. On the other hand, coherence is gradually lost for $\mathrm{m}_{\phi}\!\gtrsim\!10^{-2}\!$ eV. The sensitivity of helioscopes in terms of the axion-photon coupling strength is approximately \cite{CAST:2003idc}

\begin{equation}
\mathrm{g_{\phi \gamma}}[\mathrm{GeV^{-1}}]\!\gtrsim\!\!1.4\times10^{-9}  \!\!\frac{b^{\nicefrac{1}{8}}}{t^{\nicefrac{1}{8}}\!\left(B\ell[\mathrm{T\!\!\cdot\! m}]\right)^{\!\nicefrac{1}{2}}\!\left(\mathrm{A}[\mathrm{cm}^2]\right)^{\!\nicefrac{1}{4}}} \!\times\!\left(\!\frac{\Phi_{\!\phi}}{\Phi_{\!\!_{\odot}}}\!\!\right)^{\!\!\!\!\nicefrac{-\!1}{2}}\!\!\!\!\!,
\label{Eq.9}
\end{equation}
where $b$ is the integrated background noise in the energy range 1-10 keV, $t$ is exposure time, $\mathrm{A}$ is the transverse area of the detector and $\Phi_{\!\phi}$ is the received flux in units of $\Phi_{\!_{\!\odot}}$, the integrated flux of solar axions on Earth.

Helioscope-type apparatus can be employed to search for axion flares. In Fig.$\,$\ref{Fig.4}, we qualitatively compare solar and magnetar scanning performed by an IAXO-like instrument \cite{Irastorza:2011gs, IAXO:2019mpb}. There, we find a volume of special interest in a $\lesssim$3 kpc radius from Earth, with around three dozen cataloged magnetars, being about $\nicefrac{1}{3}$ of them within a $\nicefrac{1}{2}$ kpc radius; while the closest object is  about 0.1 kpc distant \cite{10.1093/mnras/stx2679}.

\begin{figure}[b]\centering
\includegraphics[width=.38\textwidth]{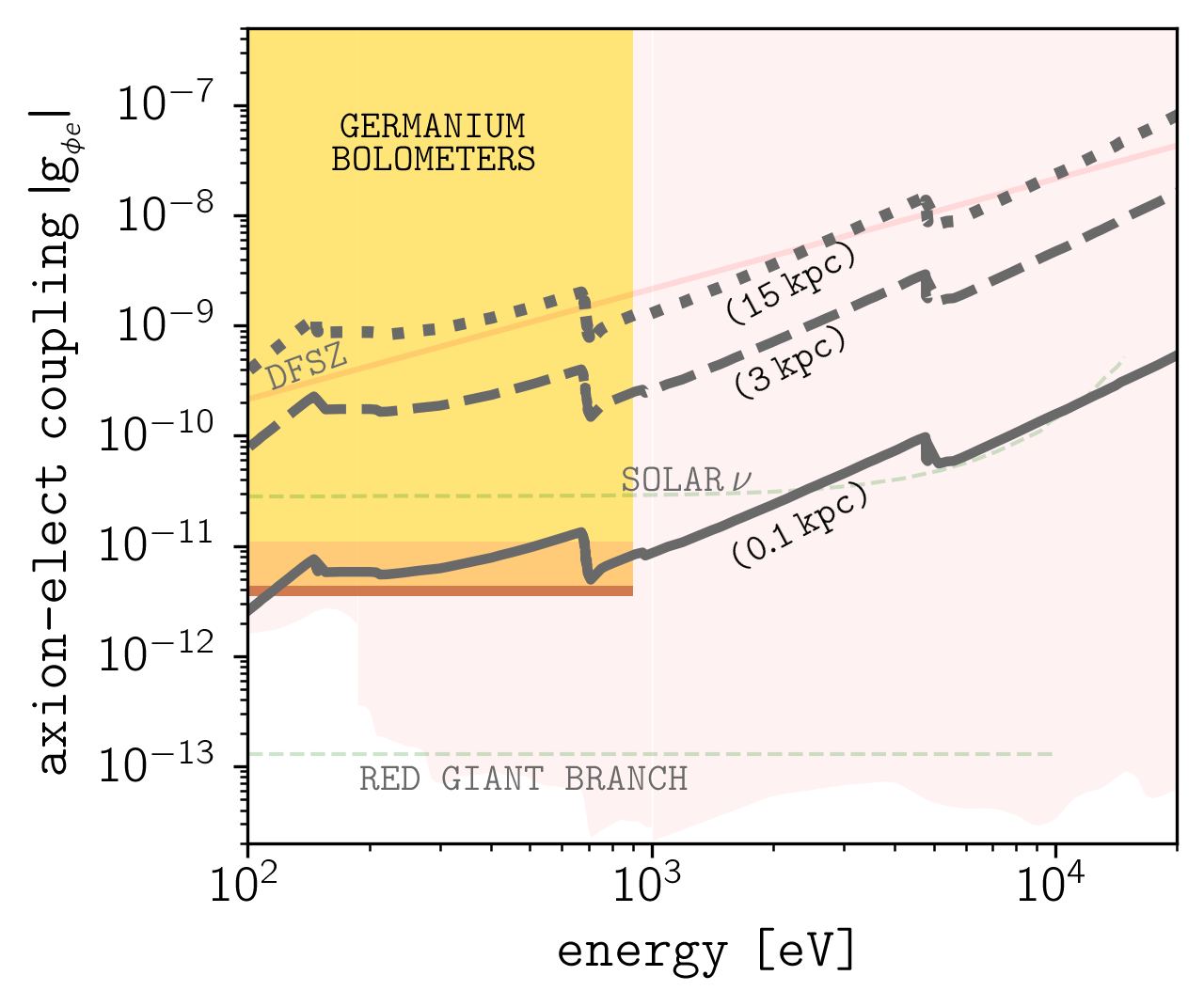}
\caption{Limit on the sensitivity to axion giant flares (AGFs) based on the axio-electric effect for realistic liquid-xenon detectors. The regions shaded in warm colors are established by direct detection of solar axion at 90\% confidence level (CL) \cite{EDELWEISS:2018tde, LUX:2017glr, PandaX:2017ock}. Dashed light-green lines project stellar hints \cite{Gondolo:2008dd, Capozzi:2020cbu,Straniero:2020iyi}. The light reddish zone corresponds with methods whose hypothesize that all the dark matter in the Halo is formed of axion and a high homogeneity, overlapping other model-dependent, or indirect, results \cite{EDELWEISS:2018tde, LUX:2017glr, PandaX:2017ock, ciaran_o_hare_2020_3932430, diluzio2021stellar}. Dark lines correspond with direct detection of giant flares with 15 — dotted —, 3 — dashed — and 0.1 — solid — kpc distance, at 95\% CL. The red solid line corresponds with DFSZ, non-relativistic, axion.  Parameters are $t\!=\!0.125$ s, W=$10^4$ kg, $\Re\!=\!0.3137$ $\sqrt{\mathrm{ke\!V}}$, $b=1.4\!\times\!10
^{-5}$ $\!$s$^{-1}$ $\!$kg$^{-1}$ $\!$keV$^{-1}$.}
\label{Fig.3}
\end{figure}

\subsection{Observation of axion flares using liquid-xenon detectors}
Experiments storing liquid-xenon are sensitive to axion through the axio-electric effect — $\phi + e + Z \!\rightarrow e'+ Z$ — hold by Eq.$\,$\ref{Eq.1}. In the relativistic limit, the cross-section reads
\begin{equation}
\mathrm{\sigma}_{\!\phi e}[\mathrm{cm^2 kg^{-1}}]\simeq 10^2 \times \mathrm{g}^2_{\phi e}\! \left(\!\frac{\mathrm{E}}{ \mathrm{ke\!V}}\!\right)^{\!\!2}\!\!\times\!\left(\!\frac{\sigma_{\mathrm{\!p}e}(\mathrm{E})}{10^6\mathrm{\,b\,atom^{-1}}}\!\right)  \,,
\label{Eq.7}
\end{equation}
where the proton-electron cross-section $\sigma_{\mathrm{\!p}e}$(E) can be, e.g., interpolated from \cite{VEIGELE197351}. By the introduction of an expectation signal-to-noise ratio (SNR) in Eq.$\,$\ref{Eq.7}, it is possible to write the sensitivity of liquid-xenon detectors in terms of the axion-electron coupling strength \cite{Arisaka:2012pb}

\begin{equation}
\mathrm{g}_{\phi e} \!\gtrsim \!10^{-1} \!\! \left(\!\frac{ \mathrm{ke\!V}}{\mathrm{E}}\!\right)\!\times\!\left(\!\!\frac{10^6\mathrm{\,b\,atom^{\!-1}}}{\sigma_{\mathrm{\!p}e}(\mathrm{E})}\!\right)^{\!\!\!\nicefrac{1}{2}} \!\!\!\!\!\times\! \sqrt{\mathrm{SNR}} \! \left(\!\!\frac{4\Re b\sqrt{\mathrm{E}}}{t\!\times\!\mathrm{W}}\!\right)^{\!\!\!\nicefrac{1}{4}}\!\!\!\!\!\times \Phi_{\!\!\phi}^{\!\!\nicefrac{-\!1}{2}} \!\!,
\label{Eq.8}
\end{equation}
where $\Re$ is the spectral resolution of the instrument, $b$ is the background event rate, $t$ is integration time, $\mathrm{W}$ is weight of the stored xenon. In Fig.$\,$\ref{Fig.3} we represent the sensitivity to AGFs for XAX-like experiments \cite{ARISAKA200963}.

\section{Conclusions}
The conversion of the beam of light emitted from highly magnetic neutron stars into relativistic axions, first at their magnetospheres and then during their flight to Earth, opens a new window for the direct detection of axion, or for re-visiting and extending exclusion limits whose arbitrarily assume that axion-like particles form all the dark matter (DM) in the nearby universe, and that DM is distributed homogeneously. In this manuscript we pioneered the concept of photon-axion multimessenger astronomy with giant flares (GFs), a type of rare event featured by soft gamma-ray repeaters (SGRs) that can be, for a fraction of a second, more luminous than hundred times the brightest supernovae \cite{Dong:2015oea, Kankare:2017rtx}, or almost $10^{14}$ suns radiating coherently; with an uncertain upper bound, as more magnetic magnetars could release GFs 1--2 orders brighter than the strongest event observed to date if we only take into account stored magnetic energy \cite{2005Natur.434.1098H}.

Giant flares could correlate with an axion counterpart via resonant mixing through the magnetosphere. Anticipating a short-term at which the SGR catalogue is more extensive and their characteristics better understood, we analyze sensitivities to axion bursts in the soft X-ray energy range for realistic experiments. Under a number of idealizations of recurring use in related works, we find that GFs originating in the vicinity of Earth would provide the detectors with sensitivity to axion-like particles with an axion-photon coupling $\mathrm{g}_{\phi \gamma}\!\gtrsim\!\mathrm{several}\!\times\!10^{-13}$ GeV$^{-1}$ and axion-electron coupling $\mathrm{g}_{\phi e}\!\gtrsim\!\mathrm{few}\!\times\!10^{-12}$ at 95\% confidence level (CL), over a broad range and for reasonable benchmark parameters. A confirmed detection would result in an absolute `spectra', while for the projection of new exclusion bounds the sensitivity would be calibrated against data from X-ray telescopes and grounded by theory. The expected axion flux density on Earth results from the convolution of three terms: the GF photon spectral flux density, the photon-to-axion conversion probability and the geometric dilution with distance. As a reference, for an upper-limit photon-axion conversion probability $\pazocal{O}(\nicefrac{1}{2})$, the flare from SGR 1806-20 — $10^{46}$erg, 15 kpc distance — would result in a transient fluence on Earth similar to the flux of solar axions for an axion-photon coupling strength $\mathrm{g}_{\phi\gamma}\!\!\sim\!\! 10^{-11}$ GeV$^{-1}$. 

Assuming axion, this work predicts the existence of axion giant flares (AGFs) across our universe, and perhaps their detectability by ground-based observatories. Transcendentally, the echo of a galactic flare might be received today in the form of AGF, due to the delay-time that massive particles suffer compared to ordinary photons during their flight through the interstellar medium. Thus, monitoring the confirmed sources of giant flares — SGR 0526-66, 48 kpc distant, in 1979; SGR 1900+14, 6 kpc distant, in 1998; and SGR 1806-20 in 2004 — with dark matter detectors, in addition to future events, can be motivated — e.g., during hours when helioscopes cannot track the Sun —. A state-of-the-art estimate of the event rate suggests that each magnetar releases up to one potent giant flare every fifty years \cite{Burns_2021}. There are currently more than four dozen cataloged magnetars within a radius of 15 kpc from Earth. Therefore, the probability of an observable AGF event would be in the order of a dozen in the next decade, with approximately 2/3 of them originating within a distance of 3 kpc. In addition, it is striking that, although the bursts themselves are not periodic, the activity might only occur during predictable periodic intervals \cite{PhysRevD.104.023007}. As a consequence, the prediction and tracking of magnetars entering an active period using  dedicated axio-telescopes, or a network, is not discardable in order to enhance the probability of detection through the observation of stars with active, non-overlapping windows.

Finally, the XENON1T Collaboration recently reported an electronic recoil excess below 7 keV compatible with solar axion at 3.4$\sigma$ CL \cite{XENON:2020rca}. However, the precise parameter space is in tension with stellar evolution at 8$\sigma$ \cite{DiLuzio:2020jjp}. Interestingly, magnetar axions could mimic solar axion at keV energies without conflicting with stellar physics. However, the signal persisted for a large time interval incompatible with the known nature of magnetar flares, while quiescent isolated sources would be too distant to provide the deposited energy.
Magnetar axion count rate could be a factor to consider in future experiments.

\section*{Acknowledgements}
This work was supported by the Special Postdoctoral Researchers (SPDR) program. JDM appreciates having been invited Visiting Researcher of the Instituto de Astrofísica de Canarias (IAC) for a period coinciding with a part of this research.
\bibliography{apssamp}

\begin{thebibliography}{70}%
\makeatletter
\providecommand \@ifxundefined [1]{%
 \@ifx{#1\undefined}
}%
\providecommand \@ifnum [1]{%
 \ifnum #1\expandafter \@firstoftwo
 \else \expandafter \@secondoftwo
 \fi
}%
\providecommand \@ifx [1]{%
 \ifx #1\expandafter \@firstoftwo
 \else \expandafter \@secondoftwo
 \fi
}%
\providecommand \natexlab [1]{#1}%
\providecommand \enquote  [1]{``#1''}%
\providecommand \bibnamefont  [1]{#1}%
\providecommand \bibfnamefont [1]{#1}%
\providecommand \citenamefont [1]{#1}%
\providecommand \href@noop [0]{\@secondoftwo}%
\providecommand \href [0]{\begingroup \@sanitize@url \@href}%
\providecommand \@href[1]{\@@startlink{#1}\@@href}%
\providecommand \@@href[1]{\endgroup#1\@@endlink}%
\providecommand \@sanitize@url [0]{\catcode `\\12\catcode `\$12\catcode
  `\&12\catcode `\#12\catcode `\^12\catcode `\_12\catcode `\%12\relax}%
\providecommand \@@startlink[1]{}%
\providecommand \@@endlink[0]{}%
\providecommand \url  [0]{\begingroup\@sanitize@url \@url }%
\providecommand \@url [1]{\endgroup\@href {#1}{\urlprefix }}%
\providecommand \urlprefix  [0]{URL }%
\providecommand \Eprint [0]{\href }%
\providecommand \doibase [0]{https://doi.org/}%
\providecommand \selectlanguage [0]{\@gobble}%
\providecommand \bibinfo  [0]{\@secondoftwo}%
\providecommand \bibfield  [0]{\@secondoftwo}%
\providecommand \translation [1]{[#1]}%
\providecommand \BibitemOpen [0]{}%
\providecommand \bibitemStop [0]{}%
\providecommand \bibitemNoStop [0]{.\EOS\space}%
\providecommand \EOS [0]{\spacefactor3000\relax}%
\providecommand \BibitemShut  [1]{\csname bibitem#1\endcsname}%
\let\auto@bib@innerbib\@empty
\bibitem [{\citenamefont {Weinberg}(1978)}]{PhysRevLett.40.223}%
  \BibitemOpen
  \bibfield  {author} {\bibinfo {author} {\bibfnamefont {S.}~\bibnamefont
  {Weinberg}},\ }\bibfield  {title} {\bibinfo {title} {A new light boson?},\
  }\href {https://doi.org/10.1103/PhysRevLett.40.223} {\bibfield  {journal}
  {\bibinfo  {journal} {Phys. Rev. Lett.}\ }\textbf {\bibinfo {volume} {40}},\
  \bibinfo {pages} {223} (\bibinfo {year} {1978})}\BibitemShut {NoStop}%
\bibitem [{\citenamefont {Wilczek}(1978)}]{PhysRevLett.40.279}%
  \BibitemOpen
  \bibfield  {author} {\bibinfo {author} {\bibfnamefont {F.}~\bibnamefont
  {Wilczek}},\ }\bibfield  {title} {\bibinfo {title} {Problem of strong $p$ and
  $t$ invariance in the presence of instantons},\ }\href
  {https://doi.org/10.1103/PhysRevLett.40.279} {\bibfield  {journal} {\bibinfo
  {journal} {Phys. Rev. Lett.}\ }\textbf {\bibinfo {volume} {40}},\ \bibinfo
  {pages} {279} (\bibinfo {year} {1978})}\BibitemShut {NoStop}%
\bibitem [{\citenamefont {Peccei}\ and\ \citenamefont
  {Quinn}(1977)}]{PhysRevLett.38.1440}%
  \BibitemOpen
  \bibfield  {author} {\bibinfo {author} {\bibfnamefont {R.~D.}\ \bibnamefont
  {Peccei}}\ and\ \bibinfo {author} {\bibfnamefont {H.~R.}\ \bibnamefont
  {Quinn}},\ }\bibfield  {title} {\bibinfo {title} {$\mathrm{CP}$ conservation
  in the presence of pseudoparticles},\ }\href
  {https://doi.org/10.1103/PhysRevLett.38.1440} {\bibfield  {journal} {\bibinfo
   {journal} {Phys. Rev. Lett.}\ }\textbf {\bibinfo {volume} {38}},\ \bibinfo
  {pages} {1440} (\bibinfo {year} {1977})}\BibitemShut {NoStop}%
\bibitem [{\citenamefont {Abbott}\ and\ \citenamefont
  {Sikivie}(1983)}]{ABBOTT1983133}%
  \BibitemOpen
  \bibfield  {author} {\bibinfo {author} {\bibfnamefont {L.}~\bibnamefont
  {Abbott}}\ and\ \bibinfo {author} {\bibfnamefont {P.}~\bibnamefont
  {Sikivie}},\ }\bibfield  {title} {\bibinfo {title} {A cosmological bound on
  the invisible axion},\ }\href
  {https://doi.org/https://doi.org/10.1016/0370-2693(83)90638-X} {\bibfield
  {journal} {\bibinfo  {journal} {Physics Letters B}\ }\textbf {\bibinfo
  {volume} {120}},\ \bibinfo {pages} {133} (\bibinfo {year}
  {1983})}\BibitemShut {NoStop}%
\bibitem [{\citenamefont {Dine}\ and\ \citenamefont
  {Fischler}(1983)}]{DINE1983137}%
  \BibitemOpen
  \bibfield  {author} {\bibinfo {author} {\bibfnamefont {M.}~\bibnamefont
  {Dine}}\ and\ \bibinfo {author} {\bibfnamefont {W.}~\bibnamefont
  {Fischler}},\ }\bibfield  {title} {\bibinfo {title} {The not-so-harmless
  axion},\ }\href
  {https://doi.org/https://doi.org/10.1016/0370-2693(83)90639-1} {\bibfield
  {journal} {\bibinfo  {journal} {Physics Letters B}\ }\textbf {\bibinfo
  {volume} {120}},\ \bibinfo {pages} {137} (\bibinfo {year}
  {1983})}\BibitemShut {NoStop}%
\bibitem [{\citenamefont {Preskill}\ \emph {et~al.}(1983)\citenamefont
  {Preskill}, \citenamefont {Wise},\ and\ \citenamefont
  {Wilczek}}]{PRESKILL1983127}%
  \BibitemOpen
  \bibfield  {author} {\bibinfo {author} {\bibfnamefont {J.}~\bibnamefont
  {Preskill}}, \bibinfo {author} {\bibfnamefont {M.~B.}\ \bibnamefont {Wise}},\
  and\ \bibinfo {author} {\bibfnamefont {F.}~\bibnamefont {Wilczek}},\
  }\bibfield  {title} {\bibinfo {title} {Cosmology of the invisible axion},\
  }\href {https://doi.org/https://doi.org/10.1016/0370-2693(83)90637-8}
  {\bibfield  {journal} {\bibinfo  {journal} {Physics Letters B}\ }\textbf
  {\bibinfo {volume} {120}},\ \bibinfo {pages} {127} (\bibinfo {year}
  {1983})}\BibitemShut {NoStop}%
\bibitem [{\citenamefont {{Particle Data Group}}(2018)}]{PhysRevD.98.030001}%
  \BibitemOpen
  \bibfield  {author} {\bibinfo {author} {\bibnamefont {{Particle Data
  Group}}},\ }\bibfield  {title} {\bibinfo {title} {Review of particle
  physics},\ }\href {https://doi.org/10.1103/PhysRevD.98.030001} {\bibfield
  {journal} {\bibinfo  {journal} {Phys. Rev. D}\ }\textbf {\bibinfo {volume}
  {98}},\ \bibinfo {pages} {030001} (\bibinfo {year} {2018})}\BibitemShut
  {NoStop}%
\bibitem [{\citenamefont {Wilczek}(1987)}]{Wilczek:1987mv}%
  \BibitemOpen
  \bibfield  {author} {\bibinfo {author} {\bibfnamefont {F.}~\bibnamefont
  {Wilczek}},\ }\bibfield  {title} {\bibinfo {title} {{Two Applications of
  Axion Electrodynamics}},\ }\href
  {https://doi.org/10.1103/PhysRevLett.58.1799} {\bibfield  {journal} {\bibinfo
   {journal} {Phys. Rev. Lett.}\ }\textbf {\bibinfo {volume} {58}},\ \bibinfo
  {pages} {1799} (\bibinfo {year} {1987})}\BibitemShut {NoStop}%
\bibitem [{\citenamefont {Raffelt}\ and\ \citenamefont
  {Stodolsky}(1988)}]{PhysRevD.37.1237}%
  \BibitemOpen
  \bibfield  {author} {\bibinfo {author} {\bibfnamefont {G.}~\bibnamefont
  {Raffelt}}\ and\ \bibinfo {author} {\bibfnamefont {L.}~\bibnamefont
  {Stodolsky}},\ }\bibfield  {title} {\bibinfo {title} {Mixing of the photon
  with low-mass particles},\ }\href {https://doi.org/10.1103/PhysRevD.37.1237}
  {\bibfield  {journal} {\bibinfo  {journal} {Phys. Rev. D}\ }\textbf {\bibinfo
  {volume} {37}},\ \bibinfo {pages} {1237} (\bibinfo {year}
  {1988})}\BibitemShut {NoStop}%
\bibitem [{\citenamefont {{Goldreich}}\ and\ \citenamefont
  {{Julian}}(1969)}]{1969ApJ...157..869G}%
  \BibitemOpen
  \bibfield  {author} {\bibinfo {author} {\bibfnamefont {P.}~\bibnamefont
  {{Goldreich}}}\ and\ \bibinfo {author} {\bibfnamefont {W.~H.}\ \bibnamefont
  {{Julian}}},\ }\bibfield  {title} {\bibinfo {title} {{Pulsar
  Electrodynamics}},\ }\href {https://doi.org/10.1086/150119} {\bibfield
  {journal} {\bibinfo  {journal} {\apj}\ }\textbf {\bibinfo {volume} {157}},\
  \bibinfo {pages} {869} (\bibinfo {year} {1969})}\BibitemShut {NoStop}%
\bibitem [{\citenamefont {{De Miguel}}\ and\ \citenamefont
  {Otani}(2021)}]{demiguel2021superdense}%
  \BibitemOpen
  \bibfield  {author} {\bibinfo {author} {\bibfnamefont {J.}~\bibnamefont {{De
  Miguel}}}\ and\ \bibinfo {author} {\bibfnamefont {C.}~\bibnamefont {Otani}},\
  }\href@noop {} {\bibinfo {title} {Superdense beaming of axion dark matter in
  the vicinity of the light cylinder of pulsars}} (\bibinfo {year} {2021}),\
  \Eprint {https://arxiv.org/abs/2111.01746} {arXiv:2111.01746 [astro-ph.HE]}
  \BibitemShut {NoStop}%
\bibitem [{\citenamefont {Huang}\ \emph {et~al.}(2018)\citenamefont {Huang},
  \citenamefont {Kadota}, \citenamefont {Sekiguchi},\ and\ \citenamefont
  {Tashiro}}]{PhysRevD.97.123001}%
  \BibitemOpen
  \bibfield  {author} {\bibinfo {author} {\bibfnamefont {F.~P.}\ \bibnamefont
  {Huang}}, \bibinfo {author} {\bibfnamefont {K.}~\bibnamefont {Kadota}},
  \bibinfo {author} {\bibfnamefont {T.}~\bibnamefont {Sekiguchi}},\ and\
  \bibinfo {author} {\bibfnamefont {H.}~\bibnamefont {Tashiro}},\ }\bibfield
  {title} {\bibinfo {title} {Radio telescope search for the resonant conversion
  of cold dark matter axions from the magnetized astrophysical sources},\
  }\href {https://doi.org/10.1103/PhysRevD.97.123001} {\bibfield  {journal}
  {\bibinfo  {journal} {Phys. Rev. D}\ }\textbf {\bibinfo {volume} {97}},\
  \bibinfo {pages} {123001} (\bibinfo {year} {2018})}\BibitemShut {NoStop}%
\bibitem [{\citenamefont {Roberts}\ \emph {et~al.}(2021)\citenamefont {Roberts}
  \emph {et~al.}}]{Roberts:2021udn}%
  \BibitemOpen
  \bibfield  {author} {\bibinfo {author} {\bibfnamefont {O.~J.}\ \bibnamefont
  {Roberts}} \emph {et~al.},\ }\bibfield  {title} {\bibinfo {title} {{Rapid
  spectral variability of a giant flare from a magnetar in NGC 253}},\ }\href
  {https://doi.org/10.1038/s41586-020-03077-8} {\bibfield  {journal} {\bibinfo
  {journal} {Nature}\ }\textbf {\bibinfo {volume} {589}},\ \bibinfo {pages}
  {207} (\bibinfo {year} {2021})},\ \Eprint {https://arxiv.org/abs/2101.05146}
  {arXiv:2101.05146 [astro-ph.HE]} \BibitemShut {NoStop}%
\bibitem [{\citenamefont {{Yang}}\ and\ \citenamefont
  {{Zhang}}(2015)}]{2015ApJ...815...45Y}%
  \BibitemOpen
  \bibfield  {author} {\bibinfo {author} {\bibfnamefont {Y.-P.}\ \bibnamefont
  {{Yang}}}\ and\ \bibinfo {author} {\bibfnamefont {B.}~\bibnamefont
  {{Zhang}}},\ }\bibfield  {title} {\bibinfo {title} {{On the Polarization
  Properties of Magnetar Giant Flare Pulsating Tails}},\ }\href
  {https://doi.org/10.1088/0004-637X/815/1/45} {\bibfield  {journal} {\bibinfo
  {journal} {\apj}\ }\textbf {\bibinfo {volume} {815}},\ \bibinfo {eid} {45}
  (\bibinfo {year} {2015})},\ \Eprint {https://arxiv.org/abs/1511.00475}
  {arXiv:1511.00475 [astro-ph.HE]} \BibitemShut {NoStop}%
\bibitem [{\citenamefont {Gu\'epin}\ \emph {et~al.}(2020)\citenamefont
  {Gu\'epin}, \citenamefont {Cerutti},\ and\ \citenamefont
  {Kotera}}]{Guepin:2019fjb}%
  \BibitemOpen
  \bibfield  {author} {\bibinfo {author} {\bibfnamefont {C.}~\bibnamefont
  {Gu\'epin}}, \bibinfo {author} {\bibfnamefont {B.}~\bibnamefont {Cerutti}},\
  and\ \bibinfo {author} {\bibfnamefont {K.}~\bibnamefont {Kotera}},\
  }\bibfield  {title} {\bibinfo {title} {{Proton acceleration in pulsar
  magnetospheres}},\ }\href {https://doi.org/10.1051/0004-6361/201936816}
  {\bibfield  {journal} {\bibinfo  {journal} {Astron. Astrophys.}\ }\textbf
  {\bibinfo {volume} {635}},\ \bibinfo {pages} {A138} (\bibinfo {year}
  {2020})},\ \Eprint {https://arxiv.org/abs/1910.11387} {arXiv:1910.11387
  [astro-ph.HE]} \BibitemShut {NoStop}%
\bibitem [{\citenamefont {Duncan}\ and\ \citenamefont
  {Thompson}(1992)}]{Duncan1992FormationOV}%
  \BibitemOpen
  \bibfield  {author} {\bibinfo {author} {\bibfnamefont {R.~C.}\ \bibnamefont
  {Duncan}}\ and\ \bibinfo {author} {\bibfnamefont {C.}~\bibnamefont
  {Thompson}},\ }\bibfield  {title} {\bibinfo {title} {Formation of very
  strongly magnetized neutron stars - implications for gamma-ray bursts},\
  }\href@noop {} {\bibfield  {journal} {\bibinfo  {journal} {The Astrophysical
  Journal}\ }\textbf {\bibinfo {volume} {392}} (\bibinfo {year}
  {1992})}\BibitemShut {NoStop}%
\bibitem [{\citenamefont {{Ruderman}}\ and\ \citenamefont
  {{Sutherland}}(1975)}]{1975ApJ...196...51R}%
  \BibitemOpen
  \bibfield  {author} {\bibinfo {author} {\bibfnamefont {M.~A.}\ \bibnamefont
  {{Ruderman}}}\ and\ \bibinfo {author} {\bibfnamefont {P.~G.}\ \bibnamefont
  {{Sutherland}}},\ }\bibfield  {title} {\bibinfo {title} {{Theory of pulsars:
  polar gaps, sparks, and coherent microwave radiation.}},\ }\href
  {https://doi.org/10.1086/153393} {\bibfield  {journal} {\bibinfo  {journal}
  {\apj}\ }\textbf {\bibinfo {volume} {196}},\ \bibinfo {pages} {51} (\bibinfo
  {year} {1975})}\BibitemShut {NoStop}%
\bibitem [{\citenamefont {{Arons}}\ and\ \citenamefont
  {{Scharlemann}}(1979)}]{1979ApJ...231..854A}%
  \BibitemOpen
  \bibfield  {author} {\bibinfo {author} {\bibfnamefont {J.}~\bibnamefont
  {{Arons}}}\ and\ \bibinfo {author} {\bibfnamefont {E.~T.}\ \bibnamefont
  {{Scharlemann}}},\ }\bibfield  {title} {\bibinfo {title} {{Pair formation
  above pulsar polar caps: structure of the low altitude acceleration zone.}},\
  }\href {https://doi.org/10.1086/157250} {\bibfield  {journal} {\bibinfo
  {journal} {\apj}\ }\textbf {\bibinfo {volume} {231}},\ \bibinfo {pages} {854}
  (\bibinfo {year} {1979})}\BibitemShut {NoStop}%
\bibitem [{\citenamefont {Lai}\ and\ \citenamefont
  {Heyl}(2006)}]{PhysRevD.74.123003}%
  \BibitemOpen
  \bibfield  {author} {\bibinfo {author} {\bibfnamefont {D.}~\bibnamefont
  {Lai}}\ and\ \bibinfo {author} {\bibfnamefont {J.}~\bibnamefont {Heyl}},\
  }\bibfield  {title} {\bibinfo {title} {Probing axions with radiation from
  magnetic stars},\ }\href {https://doi.org/10.1103/PhysRevD.74.123003}
  {\bibfield  {journal} {\bibinfo  {journal} {Phys. Rev. D}\ }\textbf {\bibinfo
  {volume} {74}},\ \bibinfo {pages} {123003} (\bibinfo {year}
  {2006})}\BibitemShut {NoStop}%
\bibitem [{\citenamefont {Perna}\ \emph {et~al.}(2012)\citenamefont {Perna},
  \citenamefont {Ho}, \citenamefont {Verde}, \citenamefont {van Adelsberg},\
  and\ \citenamefont {Jimenez}}]{Perna:2012wn}%
  \BibitemOpen
  \bibfield  {author} {\bibinfo {author} {\bibfnamefont {R.}~\bibnamefont
  {Perna}}, \bibinfo {author} {\bibfnamefont {W.~C.~G.}\ \bibnamefont {Ho}},
  \bibinfo {author} {\bibfnamefont {L.}~\bibnamefont {Verde}}, \bibinfo
  {author} {\bibfnamefont {M.}~\bibnamefont {van Adelsberg}},\ and\ \bibinfo
  {author} {\bibfnamefont {R.}~\bibnamefont {Jimenez}},\ }\bibfield  {title}
  {\bibinfo {title} {{Signatures of photon-axion conversion in the thermal
  spectra and polarization of neutron stars}},\ }\href
  {https://doi.org/10.1088/0004-637X/748/2/116} {\bibfield  {journal} {\bibinfo
   {journal} {Astrophys. J.}\ }\textbf {\bibinfo {volume} {748}},\ \bibinfo
  {pages} {116} (\bibinfo {year} {2012})},\ \Eprint
  {https://arxiv.org/abs/1201.5390} {arXiv:1201.5390 [astro-ph.HE]}
  \BibitemShut {NoStop}%
\bibitem [{\citenamefont {Battye}\ \emph {et~al.}(2020)\citenamefont {Battye},
  \citenamefont {Garbrecht}, \citenamefont {McDonald}, \citenamefont {Pace},\
  and\ \citenamefont {Srinivasan}}]{PhysRevD.102.023504}%
  \BibitemOpen
  \bibfield  {author} {\bibinfo {author} {\bibfnamefont {R.~A.}\ \bibnamefont
  {Battye}}, \bibinfo {author} {\bibfnamefont {B.}~\bibnamefont {Garbrecht}},
  \bibinfo {author} {\bibfnamefont {J.~I.}\ \bibnamefont {McDonald}}, \bibinfo
  {author} {\bibfnamefont {F.}~\bibnamefont {Pace}},\ and\ \bibinfo {author}
  {\bibfnamefont {S.}~\bibnamefont {Srinivasan}},\ }\bibfield  {title}
  {\bibinfo {title} {Dark matter axion detection in the radio/mm waveband},\
  }\href {https://doi.org/10.1103/PhysRevD.102.023504} {\bibfield  {journal}
  {\bibinfo  {journal} {Phys. Rev. D}\ }\textbf {\bibinfo {volume} {102}},\
  \bibinfo {pages} {023504} (\bibinfo {year} {2020})}\BibitemShut {NoStop}%
\bibitem [{\citenamefont {Zhuravlev}\ \emph {et~al.}(2021)\citenamefont
  {Zhuravlev}, \citenamefont {Popov},\ and\ \citenamefont
  {Pshirkov}}]{Zhuravlev:2021fvm}%
  \BibitemOpen
  \bibfield  {author} {\bibinfo {author} {\bibfnamefont {A.}~\bibnamefont
  {Zhuravlev}}, \bibinfo {author} {\bibfnamefont {S.}~\bibnamefont {Popov}},\
  and\ \bibinfo {author} {\bibfnamefont {M.}~\bibnamefont {Pshirkov}},\
  }\bibfield  {title} {\bibinfo {title} {{Photon-axion mixing in thermal
  emission of isolated neutron stars}},\ }\href
  {https://doi.org/10.1016/j.physletb.2021.136615} {\bibfield  {journal}
  {\bibinfo  {journal} {Phys. Lett. B}\ }\textbf {\bibinfo {volume} {821}},\
  \bibinfo {pages} {136615} (\bibinfo {year} {2021})},\ \Eprint
  {https://arxiv.org/abs/2109.04077} {arXiv:2109.04077 [astro-ph.HE]}
  \BibitemShut {NoStop}%
\bibitem [{\citenamefont {{Olive}}\ \emph {et~al.}(2004)\citenamefont
  {{Olive}}, \citenamefont {{Hurley}}, \citenamefont {{Sakamoto}},
  \citenamefont {{Atteia}}, \citenamefont {{Crew}}, \citenamefont {{Ricker}},
  \citenamefont {{Pizzichini}}, \citenamefont {{Barraud}},\ and\ \citenamefont
  {{Kawai}}}]{2004ApJ...616.1148O}%
  \BibitemOpen
  \bibfield  {author} {\bibinfo {author} {\bibfnamefont {J.~F.}\ \bibnamefont
  {{Olive}}}, \bibinfo {author} {\bibfnamefont {K.}~\bibnamefont {{Hurley}}},
  \bibinfo {author} {\bibfnamefont {T.}~\bibnamefont {{Sakamoto}}}, \bibinfo
  {author} {\bibfnamefont {J.~L.}\ \bibnamefont {{Atteia}}}, \bibinfo {author}
  {\bibfnamefont {G.}~\bibnamefont {{Crew}}}, \bibinfo {author} {\bibfnamefont
  {G.}~\bibnamefont {{Ricker}}}, \bibinfo {author} {\bibfnamefont
  {G.}~\bibnamefont {{Pizzichini}}}, \bibinfo {author} {\bibfnamefont
  {C.}~\bibnamefont {{Barraud}}},\ and\ \bibinfo {author} {\bibfnamefont
  {N.}~\bibnamefont {{Kawai}}},\ }\bibfield  {title} {\bibinfo {title}
  {{Time-resolved X-Ray Spectral Modeling of an Intermediate Burst from SGR
  1900+14 Observed by HETE-2 FREGATE and WXM}},\ }\href
  {https://doi.org/10.1086/424957} {\bibfield  {journal} {\bibinfo  {journal}
  {\apj}\ }\textbf {\bibinfo {volume} {616}},\ \bibinfo {pages} {1148}
  (\bibinfo {year} {2004})},\ \Eprint {https://arxiv.org/abs/astro-ph/0403162}
  {arXiv:astro-ph/0403162 [astro-ph]} \BibitemShut {NoStop}%
\bibitem [{\citenamefont {Boggs}\ \emph {et~al.}(2007)\citenamefont {Boggs},
  \citenamefont {Zoglauer}, \citenamefont {Bellm}, \citenamefont {Hurley},
  \citenamefont {Lin}, \citenamefont {Smith},\ and\ \citenamefont
  {Wigger}}]{Boggs:2006uk}%
  \BibitemOpen
  \bibfield  {author} {\bibinfo {author} {\bibfnamefont {S.~E.}\ \bibnamefont
  {Boggs}}, \bibinfo {author} {\bibfnamefont {A.}~\bibnamefont {Zoglauer}},
  \bibinfo {author} {\bibfnamefont {E.}~\bibnamefont {Bellm}}, \bibinfo
  {author} {\bibfnamefont {K.}~\bibnamefont {Hurley}}, \bibinfo {author}
  {\bibfnamefont {R.~P.}\ \bibnamefont {Lin}}, \bibinfo {author} {\bibfnamefont
  {D.~M.}\ \bibnamefont {Smith}},\ and\ \bibinfo {author} {\bibfnamefont
  {C.}~\bibnamefont {Wigger}},\ }\bibfield  {title} {\bibinfo {title} {{The
  Giant Flare of December 27, 2004 from SGR 1806-20}},\ }\href
  {https://doi.org/10.1086/516732} {\bibfield  {journal} {\bibinfo  {journal}
  {Astrophys. J.}\ }\textbf {\bibinfo {volume} {661}},\ \bibinfo {pages} {458}
  (\bibinfo {year} {2007})},\ \Eprint {https://arxiv.org/abs/astro-ph/0611318}
  {arXiv:astro-ph/0611318} \BibitemShut {NoStop}%
\bibitem [{\citenamefont {Mazets}\ \emph {et~al.}(1979)\citenamefont {Mazets},
  \citenamefont {Golenetskii}, \citenamefont {Il’inskii}, \citenamefont
  {Aptekar},\ and\ \citenamefont {Guryan}}]{Mazets1979ObservationsOA}%
  \BibitemOpen
  \bibfield  {author} {\bibinfo {author} {\bibfnamefont {E.~P.}\ \bibnamefont
  {Mazets}}, \bibinfo {author} {\bibfnamefont {S.}~\bibnamefont {Golenetskii}},
  \bibinfo {author} {\bibfnamefont {V.~N.}\ \bibnamefont {Il’inskii}},
  \bibinfo {author} {\bibfnamefont {R.~L.}\ \bibnamefont {Aptekar}},\ and\
  \bibinfo {author} {\bibfnamefont {Y.~A.}\ \bibnamefont {Guryan}},\ }\bibfield
   {title} {\bibinfo {title} {Observations of a flaring x-ray pulsar in
  dorado},\ }\href@noop {} {\bibfield  {journal} {\bibinfo  {journal} {Nature}\
  }\textbf {\bibinfo {volume} {282}},\ \bibinfo {pages} {587} (\bibinfo {year}
  {1979})}\BibitemShut {NoStop}%
\bibitem [{\citenamefont {Hurley}\ \emph {et~al.}(1999)\citenamefont {Hurley}
  \emph {et~al.}}]{Hurley:1998ks}%
  \BibitemOpen
  \bibfield  {author} {\bibinfo {author} {\bibfnamefont {K.}~\bibnamefont
  {Hurley}} \emph {et~al.},\ }\bibfield  {title} {\bibinfo {title} {{A Giant,
  periodic flare from the soft gamma repeater SGR1900+14}},\ }\href
  {https://doi.org/10.1038/16199} {\bibfield  {journal} {\bibinfo  {journal}
  {Nature}\ }\textbf {\bibinfo {volume} {397}},\ \bibinfo {pages} {41}
  (\bibinfo {year} {1999})},\ \Eprint {https://arxiv.org/abs/astro-ph/9811443}
  {arXiv:astro-ph/9811443} \BibitemShut {NoStop}%
\bibitem [{\citenamefont {{Hurley}}\ \emph {et~al.}(2005)\citenamefont
  {{Hurley}}, \citenamefont {{Boggs}}, \citenamefont {{Smith}}, \citenamefont
  {{Duncan}}, \citenamefont {{Lin}}, \citenamefont {{Zoglauer}}, \citenamefont
  {{Krucker}}, \citenamefont {{Hurford}}, \citenamefont {{Hudson}},
  \citenamefont {{Wigger}}, \citenamefont {{Hajdas}}, \citenamefont
  {{Thompson}}, \citenamefont {{Mitrofanov}}, \citenamefont {{Sanin}},
  \citenamefont {{Boynton}}, \citenamefont {{Fellows}}, \citenamefont {{von
  Kienlin}}, \citenamefont {{Lichti}}, \citenamefont {{Rau}},\ and\
  \citenamefont {{Cline}}}]{2005Natur.434.1098H}%
  \BibitemOpen
  \bibfield  {author} {\bibinfo {author} {\bibfnamefont {K.}~\bibnamefont
  {{Hurley}}}, \bibinfo {author} {\bibfnamefont {S.~E.}\ \bibnamefont
  {{Boggs}}}, \bibinfo {author} {\bibfnamefont {D.~M.}\ \bibnamefont
  {{Smith}}}, \bibinfo {author} {\bibfnamefont {R.~C.}\ \bibnamefont
  {{Duncan}}}, \bibinfo {author} {\bibfnamefont {R.}~\bibnamefont {{Lin}}},
  \bibinfo {author} {\bibfnamefont {A.}~\bibnamefont {{Zoglauer}}}, \bibinfo
  {author} {\bibfnamefont {S.}~\bibnamefont {{Krucker}}}, \bibinfo {author}
  {\bibfnamefont {G.}~\bibnamefont {{Hurford}}}, \bibinfo {author}
  {\bibfnamefont {H.}~\bibnamefont {{Hudson}}}, \bibinfo {author}
  {\bibfnamefont {C.}~\bibnamefont {{Wigger}}}, \bibinfo {author}
  {\bibfnamefont {W.}~\bibnamefont {{Hajdas}}}, \bibinfo {author}
  {\bibfnamefont {C.}~\bibnamefont {{Thompson}}}, \bibinfo {author}
  {\bibfnamefont {I.}~\bibnamefont {{Mitrofanov}}}, \bibinfo {author}
  {\bibfnamefont {A.}~\bibnamefont {{Sanin}}}, \bibinfo {author} {\bibfnamefont
  {W.}~\bibnamefont {{Boynton}}}, \bibinfo {author} {\bibfnamefont
  {C.}~\bibnamefont {{Fellows}}}, \bibinfo {author} {\bibfnamefont
  {A.}~\bibnamefont {{von Kienlin}}}, \bibinfo {author} {\bibfnamefont
  {G.}~\bibnamefont {{Lichti}}}, \bibinfo {author} {\bibfnamefont
  {A.}~\bibnamefont {{Rau}}},\ and\ \bibinfo {author} {\bibfnamefont
  {T.}~\bibnamefont {{Cline}}},\ }\bibfield  {title} {\bibinfo {title} {{An
  exceptionally bright flare from SGR 1806-20 and the origins of short-duration
  {\ensuremath{\gamma}}-ray bursts}},\ }\href
  {https://doi.org/10.1038/nature03519} {\bibfield  {journal} {\bibinfo
  {journal} {\nat}\ }\textbf {\bibinfo {volume} {434}},\ \bibinfo {pages}
  {1098} (\bibinfo {year} {2005})},\ \Eprint
  {https://arxiv.org/abs/astro-ph/0502329} {arXiv:astro-ph/0502329 [astro-ph]}
  \BibitemShut {NoStop}%
\bibitem [{\citenamefont {{Palmer}}\ \emph {et~al.}(2005)\citenamefont
  {{Palmer}}, \citenamefont {{Barthelmy}}, \citenamefont {{Gehrels}},
  \citenamefont {{Kippen}}, \citenamefont {{Cayton}}, \citenamefont
  {{Kouveliotou}}, \citenamefont {{Eichler}}, \citenamefont {{Wijers}},
  \citenamefont {{Woods}}, \citenamefont {{Granot}}, \citenamefont
  {{Lyubarsky}}, \citenamefont {{Ramirez-Ruiz}}, \citenamefont {{Barbier}},
  \citenamefont {{Chester}}, \citenamefont {{Cummings}}, \citenamefont
  {{Fenimore}}, \citenamefont {{Finger}}, \citenamefont {{Gaensler}},
  \citenamefont {{Hullinger}}, \citenamefont {{Krimm}}, \citenamefont
  {{Markwardt}}, \citenamefont {{Nousek}}, \citenamefont {{Parsons}},
  \citenamefont {{Patel}}, \citenamefont {{Sakamoto}}, \citenamefont {{Sato}},
  \citenamefont {{Suzuki}},\ and\ \citenamefont
  {{Tueller}}}]{2005Natur.434.1107P}%
  \BibitemOpen
  \bibfield  {author} {\bibinfo {author} {\bibfnamefont {D.~M.}\ \bibnamefont
  {{Palmer}}}, \bibinfo {author} {\bibfnamefont {S.}~\bibnamefont
  {{Barthelmy}}}, \bibinfo {author} {\bibfnamefont {N.}~\bibnamefont
  {{Gehrels}}}, \bibinfo {author} {\bibfnamefont {R.~M.}\ \bibnamefont
  {{Kippen}}}, \bibinfo {author} {\bibfnamefont {T.}~\bibnamefont {{Cayton}}},
  \bibinfo {author} {\bibfnamefont {C.}~\bibnamefont {{Kouveliotou}}}, \bibinfo
  {author} {\bibfnamefont {D.}~\bibnamefont {{Eichler}}}, \bibinfo {author}
  {\bibfnamefont {R.~A.~M.~J.}\ \bibnamefont {{Wijers}}}, \bibinfo {author}
  {\bibfnamefont {P.~M.}\ \bibnamefont {{Woods}}}, \bibinfo {author}
  {\bibfnamefont {J.}~\bibnamefont {{Granot}}}, \bibinfo {author}
  {\bibfnamefont {Y.~E.}\ \bibnamefont {{Lyubarsky}}}, \bibinfo {author}
  {\bibfnamefont {E.}~\bibnamefont {{Ramirez-Ruiz}}}, \bibinfo {author}
  {\bibfnamefont {L.}~\bibnamefont {{Barbier}}}, \bibinfo {author}
  {\bibfnamefont {M.}~\bibnamefont {{Chester}}}, \bibinfo {author}
  {\bibfnamefont {J.}~\bibnamefont {{Cummings}}}, \bibinfo {author}
  {\bibfnamefont {E.~E.}\ \bibnamefont {{Fenimore}}}, \bibinfo {author}
  {\bibfnamefont {M.~H.}\ \bibnamefont {{Finger}}}, \bibinfo {author}
  {\bibfnamefont {B.~M.}\ \bibnamefont {{Gaensler}}}, \bibinfo {author}
  {\bibfnamefont {D.}~\bibnamefont {{Hullinger}}}, \bibinfo {author}
  {\bibfnamefont {H.}~\bibnamefont {{Krimm}}}, \bibinfo {author} {\bibfnamefont
  {C.~B.}\ \bibnamefont {{Markwardt}}}, \bibinfo {author} {\bibfnamefont
  {J.~A.}\ \bibnamefont {{Nousek}}}, \bibinfo {author} {\bibfnamefont
  {A.}~\bibnamefont {{Parsons}}}, \bibinfo {author} {\bibfnamefont
  {S.}~\bibnamefont {{Patel}}}, \bibinfo {author} {\bibfnamefont
  {T.}~\bibnamefont {{Sakamoto}}}, \bibinfo {author} {\bibfnamefont
  {G.}~\bibnamefont {{Sato}}}, \bibinfo {author} {\bibfnamefont
  {M.}~\bibnamefont {{Suzuki}}},\ and\ \bibinfo {author} {\bibfnamefont
  {J.}~\bibnamefont {{Tueller}}},\ }\bibfield  {title} {\bibinfo {title} {{A
  giant {\ensuremath{\gamma}}-ray flare from the magnetar SGR 1806 - 20}},\
  }\href {https://doi.org/10.1038/nature03525} {\bibfield  {journal} {\bibinfo
  {journal} {\nat}\ }\textbf {\bibinfo {volume} {434}},\ \bibinfo {pages}
  {1107} (\bibinfo {year} {2005})},\ \Eprint
  {https://arxiv.org/abs/astro-ph/0503030} {arXiv:astro-ph/0503030 [astro-ph]}
  \BibitemShut {NoStop}%
\bibitem [{\citenamefont {Hurley}(2011)}]{HURLEY20111337}%
  \BibitemOpen
  \bibfield  {author} {\bibinfo {author} {\bibfnamefont {K.}~\bibnamefont
  {Hurley}},\ }\bibfield  {title} {\bibinfo {title} {The short gamma-ray burst
  – sgr giant flare connection},\ }\href
  {https://doi.org/https://doi.org/10.1016/j.asr.2010.08.036} {\bibfield
  {journal} {\bibinfo  {journal} {Advances in Space Research}\ }\textbf
  {\bibinfo {volume} {47}},\ \bibinfo {pages} {1337} (\bibinfo {year}
  {2011})},\ \bibinfo {note} {neutron Stars and Gamma Ray Bursts}\BibitemShut
  {NoStop}%
\bibitem [{\citenamefont {Burns}\ \emph {et~al.}(2021)\citenamefont {Burns}
  \emph {et~al.}}]{Burns_2021}%
  \BibitemOpen
  \bibfield  {author} {\bibinfo {author} {\bibfnamefont {E.}~\bibnamefont
  {Burns}} \emph {et~al.},\ }\bibfield  {title} {\bibinfo {title}
  {Identification of a local sample of gamma-ray bursts consistent with a
  magnetar giant flare origin},\ }\href
  {https://doi.org/10.3847/2041-8213/abd8c8} {\bibfield  {journal} {\bibinfo
  {journal} {The Astrophysical Journal Letters}\ }\textbf {\bibinfo {volume}
  {907}},\ \bibinfo {pages} {L28} (\bibinfo {year} {2021})}\BibitemShut
  {NoStop}%
\bibitem [{\citenamefont {{Lyubarsky}}(2002)}]{2002MNRAS.332..199L}%
  \BibitemOpen
  \bibfield  {author} {\bibinfo {author} {\bibfnamefont {Y.~E.}\ \bibnamefont
  {{Lyubarsky}}},\ }\bibfield  {title} {\bibinfo {title} {{On the X-ray spectra
  of soft gamma repeaters}},\ }\href
  {https://doi.org/10.1046/j.1365-8711.2002.05290.x} {\bibfield  {journal}
  {\bibinfo  {journal} {MNRAS}\ }\textbf {\bibinfo {volume} {332}},\ \bibinfo
  {pages} {199} (\bibinfo {year} {2002})}\BibitemShut {NoStop}%
\bibitem [{\citenamefont {Yang}\ and\ \citenamefont
  {Zhang}(2015)}]{Yang:2015zhz}%
  \BibitemOpen
  \bibfield  {author} {\bibinfo {author} {\bibfnamefont {Y.-P.}\ \bibnamefont
  {Yang}}\ and\ \bibinfo {author} {\bibfnamefont {B.}~\bibnamefont {Zhang}},\
  }\bibfield  {title} {\bibinfo {title} {{On the polarization properties of
  magnetar giant flare pulsating tails}},\ }\href
  {https://doi.org/10.1088/0004-637X/815/1/45} {\bibfield  {journal} {\bibinfo
  {journal} {Astrophys. J.}\ }\textbf {\bibinfo {volume} {815}},\ \bibinfo
  {pages} {45} (\bibinfo {year} {2015})},\ \Eprint
  {https://arxiv.org/abs/1511.00475} {arXiv:1511.00475 [astro-ph.HE]}
  \BibitemShut {NoStop}%
\bibitem [{\citenamefont {van Putten}\ \emph {et~al.}(2016)\citenamefont {van
  Putten}, \citenamefont {Watts}, \citenamefont {Baring},\ and\ \citenamefont
  {Wijers}}]{vanPutten:2016qri}%
  \BibitemOpen
  \bibfield  {author} {\bibinfo {author} {\bibfnamefont {T.}~\bibnamefont {van
  Putten}}, \bibinfo {author} {\bibfnamefont {A.~L.}\ \bibnamefont {Watts}},
  \bibinfo {author} {\bibfnamefont {M.~G.}\ \bibnamefont {Baring}},\ and\
  \bibinfo {author} {\bibfnamefont {R.~A. M.~J.}\ \bibnamefont {Wijers}},\
  }\bibfield  {title} {\bibinfo {title} {{Radiative transfer simulations of
  magnetar flare beaming}},\ }\href {https://doi.org/10.1093/mnras/stw1279}
  {\bibfield  {journal} {\bibinfo  {journal} {Mon. Not. Roy. Astron. Soc.}\
  }\textbf {\bibinfo {volume} {461}},\ \bibinfo {pages} {877} (\bibinfo {year}
  {2016})},\ \Eprint {https://arxiv.org/abs/1605.08022} {arXiv:1605.08022
  [astro-ph.HE]} \BibitemShut {NoStop}%
\bibitem [{\citenamefont {Taverna}\ and\ \citenamefont
  {Turolla}(2017)}]{Taverna:2017ftz}%
  \BibitemOpen
  \bibfield  {author} {\bibinfo {author} {\bibfnamefont {R.}~\bibnamefont
  {Taverna}}\ and\ \bibinfo {author} {\bibfnamefont {R.}~\bibnamefont
  {Turolla}},\ }\bibfield  {title} {\bibinfo {title} {{On the spectrum and
  polarization of magnetar flare emission}},\ }\href
  {https://doi.org/10.1093/mnras/stx1086} {\bibfield  {journal} {\bibinfo
  {journal} {Mon. Not. Roy. Astron. Soc.}\ }\textbf {\bibinfo {volume} {469}},\
  \bibinfo {pages} {3610} (\bibinfo {year} {2017})},\ \Eprint
  {https://arxiv.org/abs/1705.01130} {arXiv:1705.01130 [astro-ph.HE]}
  \BibitemShut {NoStop}%
\bibitem [{\citenamefont {Krasnikov}(1996)}]{PhysRevLett.76.2633}%
  \BibitemOpen
  \bibfield  {author} {\bibinfo {author} {\bibfnamefont {S.~V.}\ \bibnamefont
  {Krasnikov}},\ }\bibfield  {title} {\bibinfo {title} {New astrophysical
  constraints on the light-pseudoscalar--photon coupling},\ }\href
  {https://doi.org/10.1103/PhysRevLett.76.2633} {\bibfield  {journal} {\bibinfo
   {journal} {Phys. Rev. Lett.}\ }\textbf {\bibinfo {volume} {76}},\ \bibinfo
  {pages} {2633} (\bibinfo {year} {1996})}\BibitemShut {NoStop}%
\bibitem [{\citenamefont {Simet}\ \emph {et~al.}(2008)\citenamefont {Simet},
  \citenamefont {Hooper},\ and\ \citenamefont {Serpico}}]{PhysRevD.77.063001}%
  \BibitemOpen
  \bibfield  {author} {\bibinfo {author} {\bibfnamefont {M.}~\bibnamefont
  {Simet}}, \bibinfo {author} {\bibfnamefont {D.}~\bibnamefont {Hooper}},\ and\
  \bibinfo {author} {\bibfnamefont {P.~D.}\ \bibnamefont {Serpico}},\
  }\bibfield  {title} {\bibinfo {title} {Milky way as a kiloparsec-scale
  axionscope},\ }\href {https://doi.org/10.1103/PhysRevD.77.063001} {\bibfield
  {journal} {\bibinfo  {journal} {Phys. Rev. D}\ }\textbf {\bibinfo {volume}
  {77}},\ \bibinfo {pages} {063001} (\bibinfo {year} {2008})}\BibitemShut
  {NoStop}%
\bibitem [{\citenamefont {Fairbairn}\ \emph {et~al.}(2011)\citenamefont
  {Fairbairn}, \citenamefont {Rashba},\ and\ \citenamefont
  {Troitsky}}]{PhysRevD.84.125019}%
  \BibitemOpen
  \bibfield  {author} {\bibinfo {author} {\bibfnamefont {M.}~\bibnamefont
  {Fairbairn}}, \bibinfo {author} {\bibfnamefont {T.}~\bibnamefont {Rashba}},\
  and\ \bibinfo {author} {\bibfnamefont {S.}~\bibnamefont {Troitsky}},\
  }\bibfield  {title} {\bibinfo {title} {Photon-axion mixing and ultra-high
  energy cosmic rays from bl lac type objects: Shining light through the
  universe},\ }\href {https://doi.org/10.1103/PhysRevD.84.125019} {\bibfield
  {journal} {\bibinfo  {journal} {Phys. Rev. D}\ }\textbf {\bibinfo {volume}
  {84}},\ \bibinfo {pages} {125019} (\bibinfo {year} {2011})}\BibitemShut
  {NoStop}%
\bibitem [{\citenamefont {van Bibber}\ \emph {et~al.}(1989)\citenamefont {van
  Bibber}, \citenamefont {McIntyre}, \citenamefont {Morris},\ and\
  \citenamefont {Raffelt}}]{PhysRevD.39.2089}%
  \BibitemOpen
  \bibfield  {author} {\bibinfo {author} {\bibfnamefont {K.}~\bibnamefont {van
  Bibber}}, \bibinfo {author} {\bibfnamefont {P.~M.}\ \bibnamefont {McIntyre}},
  \bibinfo {author} {\bibfnamefont {D.~E.}\ \bibnamefont {Morris}},\ and\
  \bibinfo {author} {\bibfnamefont {G.~G.}\ \bibnamefont {Raffelt}},\
  }\bibfield  {title} {\bibinfo {title} {Design for a practical laboratory
  detector for solar axions},\ }\href
  {https://doi.org/10.1103/PhysRevD.39.2089} {\bibfield  {journal} {\bibinfo
  {journal} {Phys. Rev. D}\ }\textbf {\bibinfo {volume} {39}},\ \bibinfo
  {pages} {2089} (\bibinfo {year} {1989})}\BibitemShut {NoStop}%
\bibitem [{\citenamefont {Anastassopoulos}\ \emph {et~al.}(2017)\citenamefont
  {Anastassopoulos} \emph {et~al.}}]{CAST:2017uph}%
  \BibitemOpen
  \bibfield  {author} {\bibinfo {author} {\bibfnamefont {V.}~\bibnamefont
  {Anastassopoulos}} \emph {et~al.} (\bibinfo {collaboration} {CAST}),\
  }\bibfield  {title} {\bibinfo {title} {{New CAST Limit on the Axion-Photon
  Interaction}},\ }\href {https://doi.org/10.1038/nphys4109} {\bibfield
  {journal} {\bibinfo  {journal} {Nature Phys.}\ }\textbf {\bibinfo {volume}
  {13}},\ \bibinfo {pages} {584} (\bibinfo {year} {2017})},\ \Eprint
  {https://arxiv.org/abs/1705.02290} {arXiv:1705.02290 [hep-ex]} \BibitemShut
  {NoStop}%
\bibitem [{\citenamefont {Ayala}\ \emph {et~al.}(2014)\citenamefont {Ayala},
  \citenamefont {Dom\'\i{}nguez}, \citenamefont {Giannotti}, \citenamefont
  {Mirizzi},\ and\ \citenamefont {Straniero}}]{Ayala:2014pea}%
  \BibitemOpen
  \bibfield  {author} {\bibinfo {author} {\bibfnamefont {A.}~\bibnamefont
  {Ayala}}, \bibinfo {author} {\bibfnamefont {I.}~\bibnamefont
  {Dom\'\i{}nguez}}, \bibinfo {author} {\bibfnamefont {M.}~\bibnamefont
  {Giannotti}}, \bibinfo {author} {\bibfnamefont {A.}~\bibnamefont {Mirizzi}},\
  and\ \bibinfo {author} {\bibfnamefont {O.}~\bibnamefont {Straniero}},\
  }\bibfield  {title} {\bibinfo {title} {{Revisiting the bound on axion-photon
  coupling from Globular Clusters}},\ }\href
  {https://doi.org/10.1103/PhysRevLett.113.191302} {\bibfield  {journal}
  {\bibinfo  {journal} {Phys. Rev. Lett.}\ }\textbf {\bibinfo {volume} {113}},\
  \bibinfo {pages} {191302} (\bibinfo {year} {2014})},\ \Eprint
  {https://arxiv.org/abs/1406.6053} {arXiv:1406.6053 [astro-ph.SR]}
  \BibitemShut {NoStop}%
\bibitem [{\citenamefont {Regis}\ \emph {et~al.}(2021)\citenamefont {Regis},
  \citenamefont {Taoso}, \citenamefont {Vaz}, \citenamefont {Brinchmann},
  \citenamefont {Zoutendijk}, \citenamefont {Bouch\'e},\ and\ \citenamefont
  {Steinmetz}}]{Regis:2020fhw}%
  \BibitemOpen
  \bibfield  {author} {\bibinfo {author} {\bibfnamefont {M.}~\bibnamefont
  {Regis}}, \bibinfo {author} {\bibfnamefont {M.}~\bibnamefont {Taoso}},
  \bibinfo {author} {\bibfnamefont {D.}~\bibnamefont {Vaz}}, \bibinfo {author}
  {\bibfnamefont {J.}~\bibnamefont {Brinchmann}}, \bibinfo {author}
  {\bibfnamefont {S.~L.}\ \bibnamefont {Zoutendijk}}, \bibinfo {author}
  {\bibfnamefont {N.~F.}\ \bibnamefont {Bouch\'e}},\ and\ \bibinfo {author}
  {\bibfnamefont {M.}~\bibnamefont {Steinmetz}},\ }\bibfield  {title} {\bibinfo
  {title} {{Searching for light in the darkness: Bounds on ALP dark matter with
  the optical MUSE-faint survey}},\ }\href
  {https://doi.org/10.1016/j.physletb.2021.136075} {\bibfield  {journal}
  {\bibinfo  {journal} {Phys. Lett. B}\ }\textbf {\bibinfo {volume} {814}},\
  \bibinfo {pages} {136075} (\bibinfo {year} {2021})},\ \Eprint
  {https://arxiv.org/abs/2009.01310} {arXiv:2009.01310 [astro-ph.CO]}
  \BibitemShut {NoStop}%
\bibitem [{\citenamefont {Grin}\ \emph {et~al.}(2007)\citenamefont {Grin},
  \citenamefont {Covone}, \citenamefont {Kneib}, \citenamefont {Kamionkowski},
  \citenamefont {Blain},\ and\ \citenamefont {Jullo}}]{Grin:2006aw}%
  \BibitemOpen
  \bibfield  {author} {\bibinfo {author} {\bibfnamefont {D.}~\bibnamefont
  {Grin}}, \bibinfo {author} {\bibfnamefont {G.}~\bibnamefont {Covone}},
  \bibinfo {author} {\bibfnamefont {J.-P.}\ \bibnamefont {Kneib}}, \bibinfo
  {author} {\bibfnamefont {M.}~\bibnamefont {Kamionkowski}}, \bibinfo {author}
  {\bibfnamefont {A.}~\bibnamefont {Blain}},\ and\ \bibinfo {author}
  {\bibfnamefont {E.}~\bibnamefont {Jullo}},\ }\bibfield  {title} {\bibinfo
  {title} {{A Telescope Search for Decaying Relic Axions}},\ }\href
  {https://doi.org/10.1103/PhysRevD.75.105018} {\bibfield  {journal} {\bibinfo
  {journal} {Phys. Rev. D}\ }\textbf {\bibinfo {volume} {75}},\ \bibinfo
  {pages} {105018} (\bibinfo {year} {2007})},\ \Eprint
  {https://arxiv.org/abs/astro-ph/0611502} {arXiv:astro-ph/0611502}
  \BibitemShut {NoStop}%
\bibitem [{\citenamefont {Irastorza}\ \emph {et~al.}(2011)\citenamefont
  {Irastorza} \emph {et~al.}}]{Irastorza:2011gs}%
  \BibitemOpen
  \bibfield  {author} {\bibinfo {author} {\bibfnamefont {I.~G.}\ \bibnamefont
  {Irastorza}} \emph {et~al.},\ }\bibfield  {title} {\bibinfo {title} {{Towards
  a new generation axion helioscope}},\ }\href
  {https://doi.org/10.1088/1475-7516/2011/06/013} {\bibfield  {journal}
  {\bibinfo  {journal} {JCAP}\ }\textbf {\bibinfo {volume} {06}},\ \bibinfo
  {pages} {013}},\ \Eprint {https://arxiv.org/abs/1103.5334} {arXiv:1103.5334
  [hep-ex]} \BibitemShut {NoStop}%
\bibitem [{\citenamefont {Kim}(1979)}]{PhysRevLett.43.103}%
  \BibitemOpen
  \bibfield  {author} {\bibinfo {author} {\bibfnamefont {J.~E.}\ \bibnamefont
  {Kim}},\ }\bibfield  {title} {\bibinfo {title} {Weak-interaction singlet and
  strong $\mathrm{CP}$ invariance},\ }\href
  {https://doi.org/10.1103/PhysRevLett.43.103} {\bibfield  {journal} {\bibinfo
  {journal} {Phys. Rev. Lett.}\ }\textbf {\bibinfo {volume} {43}},\ \bibinfo
  {pages} {103} (\bibinfo {year} {1979})}\BibitemShut {NoStop}%
\bibitem [{\citenamefont {Shifman}\ \emph {et~al.}(1980)\citenamefont
  {Shifman}, \citenamefont {Vainshtein},\ and\ \citenamefont
  {Zakharov}}]{Shifman1980CanCE}%
  \BibitemOpen
  \bibfield  {author} {\bibinfo {author} {\bibfnamefont {M.~A.}\ \bibnamefont
  {Shifman}}, \bibinfo {author} {\bibfnamefont {A.}~\bibnamefont
  {Vainshtein}},\ and\ \bibinfo {author} {\bibfnamefont {V.~I.}\ \bibnamefont
  {Zakharov}},\ }\bibfield  {title} {\bibinfo {title} {Can confinement ensure
  natural cp invariance of strong interactions},\ }\href@noop {} {\bibfield
  {journal} {\bibinfo  {journal} {Nuclear Physics}\ }\textbf {\bibinfo {volume}
  {166}},\ \bibinfo {pages} {493} (\bibinfo {year} {1980})}\BibitemShut
  {NoStop}%
\bibitem [{\citenamefont {Dine}\ \emph {et~al.}(1981)\citenamefont {Dine},
  \citenamefont {Fischler},\ and\ \citenamefont {Srednicki}}]{DINE1981199}%
  \BibitemOpen
  \bibfield  {author} {\bibinfo {author} {\bibfnamefont {M.}~\bibnamefont
  {Dine}}, \bibinfo {author} {\bibfnamefont {W.}~\bibnamefont {Fischler}},\
  and\ \bibinfo {author} {\bibfnamefont {M.}~\bibnamefont {Srednicki}},\
  }\bibfield  {title} {\bibinfo {title} {A simple solution to the strong cp
  problem with a harmless axion},\ }\href
  {https://doi.org/https://doi.org/10.1016/0370-2693(81)90590-6} {\bibfield
  {journal} {\bibinfo  {journal} {Physics Letters B}\ }\textbf {\bibinfo
  {volume} {104}},\ \bibinfo {pages} {199} (\bibinfo {year}
  {1981})}\BibitemShut {NoStop}%
\bibitem [{\citenamefont {Zhitnitskii}(1980)}]{osti_7063072}%
  \BibitemOpen
  \bibfield  {author} {\bibinfo {author} {\bibfnamefont {A.~P.}\ \bibnamefont
  {Zhitnitskii}},\ }\bibfield  {title} {\bibinfo {title} {Possible suppression
  of axion-hadron interactions},\ }\bibfield  {journal} {\bibinfo  {journal}
  {Sov. J. Nucl. Phys. (Engl. Transl.); (United States)}\ }\textbf {\bibinfo
  {volume} {31:2}},\ \href {https://www.osti.gov/biblio/7063072} {} (\bibinfo
  {year} {1980})\BibitemShut {NoStop}%
\bibitem [{\citenamefont {Sikivie}(1983)}]{PhysRevLett.51.1415}%
  \BibitemOpen
  \bibfield  {author} {\bibinfo {author} {\bibfnamefont {P.}~\bibnamefont
  {Sikivie}},\ }\bibfield  {title} {\bibinfo {title} {Experimental tests of the
  "invisible" axion},\ }\href {https://doi.org/10.1103/PhysRevLett.51.1415}
  {\bibfield  {journal} {\bibinfo  {journal} {Phys. Rev. Lett.}\ }\textbf
  {\bibinfo {volume} {51}},\ \bibinfo {pages} {1415} (\bibinfo {year}
  {1983})}\BibitemShut {NoStop}%
\bibitem [{\citenamefont {Lazarus}\ \emph {et~al.}(1992)\citenamefont
  {Lazarus}, \citenamefont {Smith}, \citenamefont {Cameron}, \citenamefont
  {Melissinos}, \citenamefont {Ruoso}, \citenamefont {Semertzidis},\ and\
  \citenamefont {Nezrick}}]{PhysRevLett.69.2333}%
  \BibitemOpen
  \bibfield  {author} {\bibinfo {author} {\bibfnamefont {D.~M.}\ \bibnamefont
  {Lazarus}}, \bibinfo {author} {\bibfnamefont {G.~C.}\ \bibnamefont {Smith}},
  \bibinfo {author} {\bibfnamefont {R.}~\bibnamefont {Cameron}}, \bibinfo
  {author} {\bibfnamefont {A.~C.}\ \bibnamefont {Melissinos}}, \bibinfo
  {author} {\bibfnamefont {G.}~\bibnamefont {Ruoso}}, \bibinfo {author}
  {\bibfnamefont {Y.~K.}\ \bibnamefont {Semertzidis}},\ and\ \bibinfo {author}
  {\bibfnamefont {F.~A.}\ \bibnamefont {Nezrick}},\ }\bibfield  {title}
  {\bibinfo {title} {Search for solar axions},\ }\href
  {https://doi.org/10.1103/PhysRevLett.69.2333} {\bibfield  {journal} {\bibinfo
   {journal} {Phys. Rev. Lett.}\ }\textbf {\bibinfo {volume} {69}},\ \bibinfo
  {pages} {2333} (\bibinfo {year} {1992})}\BibitemShut {NoStop}%
\bibitem [{\citenamefont {Moriyama}\ \emph {et~al.}(1998)\citenamefont
  {Moriyama}, \citenamefont {Minowa}, \citenamefont {Namba}, \citenamefont
  {Inoue}, \citenamefont {Takasu},\ and\ \citenamefont
  {Yamamoto}}]{Moriyama:1998kd}%
  \BibitemOpen
  \bibfield  {author} {\bibinfo {author} {\bibfnamefont {S.}~\bibnamefont
  {Moriyama}}, \bibinfo {author} {\bibfnamefont {M.}~\bibnamefont {Minowa}},
  \bibinfo {author} {\bibfnamefont {T.}~\bibnamefont {Namba}}, \bibinfo
  {author} {\bibfnamefont {Y.}~\bibnamefont {Inoue}}, \bibinfo {author}
  {\bibfnamefont {Y.}~\bibnamefont {Takasu}},\ and\ \bibinfo {author}
  {\bibfnamefont {A.}~\bibnamefont {Yamamoto}},\ }\bibfield  {title} {\bibinfo
  {title} {{Direct search for solar axions by using strong magnetic field and
  x-ray detectors}},\ }\href {https://doi.org/10.1016/S0370-2693(98)00766-7}
  {\bibfield  {journal} {\bibinfo  {journal} {Phys. Lett. B}\ }\textbf
  {\bibinfo {volume} {434}},\ \bibinfo {pages} {147} (\bibinfo {year}
  {1998})},\ \Eprint {https://arxiv.org/abs/hep-ex/9805026}
  {arXiv:hep-ex/9805026} \BibitemShut {NoStop}%
\bibitem [{\citenamefont {Zioutas}\ \emph {et~al.}(2005)\citenamefont
  {Zioutas}, \citenamefont {Andriamonje}, \citenamefont {Arsov}, \citenamefont
  {Aune}, \citenamefont {Autiero}, \citenamefont {Avignone}, \citenamefont
  {Barth}, \citenamefont {Belov}, \citenamefont {Beltr\'an}, \citenamefont
  {Br\"auninger}, \citenamefont {Carmona}, \citenamefont {Cebri\'an},
  \citenamefont {Chesi}, \citenamefont {Collar}, \citenamefont {Creswick},
  \citenamefont {Dafni}, \citenamefont {Davenport}, \citenamefont {Di~Lella},
  \citenamefont {Eleftheriadis}, \citenamefont {Englhauser}, \citenamefont
  {Fanourakis}, \citenamefont {Farach}, \citenamefont {Ferrer}, \citenamefont
  {Fischer}, \citenamefont {Franz}, \citenamefont {Friedrich}, \citenamefont
  {Geralis}, \citenamefont {Giomataris}, \citenamefont {Gninenko},
  \citenamefont {Goloubev}, \citenamefont {Hasinoff}, \citenamefont {Heinsius},
  \citenamefont {Hoffmann}, \citenamefont {Irastorza}, \citenamefont {Jacoby},
  \citenamefont {Kang}, \citenamefont {K\"onigsmann}, \citenamefont {Kotthaus},
  \citenamefont {Kr\ifmmode~\check{c}\else \v{c}\fi{}mar}, \citenamefont
  {Kousouris}, \citenamefont {Kuster}, \citenamefont
  {Laki\ifmmode~\acute{c}\else \'{c}\fi{}}, \citenamefont {Lasseur},
  \citenamefont {Liolios}, \citenamefont {Ljubi\ifmmode \check{c}\else
  \v{c}\fi{}i\ifmmode~\acute{c}\else \'{c}\fi{}}, \citenamefont {Lutz},
  \citenamefont {Luz\'on}, \citenamefont {Miller}, \citenamefont {Morales},
  \citenamefont {Morales}, \citenamefont {Mutterer}, \citenamefont
  {Nikolaidis}, \citenamefont {Ortiz}, \citenamefont {Papaevangelou},
  \citenamefont {Placci}, \citenamefont {Raffelt}, \citenamefont {Ruz},
  \citenamefont {Riege}, \citenamefont {Sarsa}, \citenamefont {Savvidis},
  \citenamefont {Serber}, \citenamefont {Serpico}, \citenamefont {Semertzidis},
  \citenamefont {Stewart}, \citenamefont {Vieira}, \citenamefont {Villar},
  \citenamefont {Walckiers},\ and\ \citenamefont
  {Zachariadou}}]{PhysRevLett.94.121301}%
  \BibitemOpen
  \bibfield  {author} {\bibinfo {author} {\bibfnamefont {K.}~\bibnamefont
  {Zioutas}}, \bibinfo {author} {\bibfnamefont {S.}~\bibnamefont
  {Andriamonje}}, \bibinfo {author} {\bibfnamefont {V.}~\bibnamefont {Arsov}},
  \bibinfo {author} {\bibfnamefont {S.}~\bibnamefont {Aune}}, \bibinfo {author}
  {\bibfnamefont {D.}~\bibnamefont {Autiero}}, \bibinfo {author} {\bibfnamefont
  {F.~T.}\ \bibnamefont {Avignone}}, \bibinfo {author} {\bibfnamefont
  {K.}~\bibnamefont {Barth}}, \bibinfo {author} {\bibfnamefont
  {A.}~\bibnamefont {Belov}}, \bibinfo {author} {\bibfnamefont
  {B.}~\bibnamefont {Beltr\'an}}, \bibinfo {author} {\bibfnamefont
  {H.}~\bibnamefont {Br\"auninger}}, \bibinfo {author} {\bibfnamefont {J.~M.}\
  \bibnamefont {Carmona}}, \bibinfo {author} {\bibfnamefont {S.}~\bibnamefont
  {Cebri\'an}}, \bibinfo {author} {\bibfnamefont {E.}~\bibnamefont {Chesi}},
  \bibinfo {author} {\bibfnamefont {J.~I.}\ \bibnamefont {Collar}}, \bibinfo
  {author} {\bibfnamefont {R.}~\bibnamefont {Creswick}}, \bibinfo {author}
  {\bibfnamefont {T.}~\bibnamefont {Dafni}}, \bibinfo {author} {\bibfnamefont
  {M.}~\bibnamefont {Davenport}}, \bibinfo {author} {\bibfnamefont
  {L.}~\bibnamefont {Di~Lella}}, \bibinfo {author} {\bibfnamefont
  {C.}~\bibnamefont {Eleftheriadis}}, \bibinfo {author} {\bibfnamefont
  {J.}~\bibnamefont {Englhauser}}, \bibinfo {author} {\bibfnamefont
  {G.}~\bibnamefont {Fanourakis}}, \bibinfo {author} {\bibfnamefont
  {H.}~\bibnamefont {Farach}}, \bibinfo {author} {\bibfnamefont
  {E.}~\bibnamefont {Ferrer}}, \bibinfo {author} {\bibfnamefont
  {H.}~\bibnamefont {Fischer}}, \bibinfo {author} {\bibfnamefont
  {J.}~\bibnamefont {Franz}}, \bibinfo {author} {\bibfnamefont
  {P.}~\bibnamefont {Friedrich}}, \bibinfo {author} {\bibfnamefont
  {T.}~\bibnamefont {Geralis}}, \bibinfo {author} {\bibfnamefont
  {I.}~\bibnamefont {Giomataris}}, \bibinfo {author} {\bibfnamefont
  {S.}~\bibnamefont {Gninenko}}, \bibinfo {author} {\bibfnamefont
  {N.}~\bibnamefont {Goloubev}}, \bibinfo {author} {\bibfnamefont {M.~D.}\
  \bibnamefont {Hasinoff}}, \bibinfo {author} {\bibfnamefont {F.~H.}\
  \bibnamefont {Heinsius}}, \bibinfo {author} {\bibfnamefont {D.~H.~H.}\
  \bibnamefont {Hoffmann}}, \bibinfo {author} {\bibfnamefont {I.~G.}\
  \bibnamefont {Irastorza}}, \bibinfo {author} {\bibfnamefont {J.}~\bibnamefont
  {Jacoby}}, \bibinfo {author} {\bibfnamefont {D.}~\bibnamefont {Kang}},
  \bibinfo {author} {\bibfnamefont {K.}~\bibnamefont {K\"onigsmann}}, \bibinfo
  {author} {\bibfnamefont {R.}~\bibnamefont {Kotthaus}}, \bibinfo {author}
  {\bibfnamefont {M.}~\bibnamefont {Kr\ifmmode~\check{c}\else \v{c}\fi{}mar}},
  \bibinfo {author} {\bibfnamefont {K.}~\bibnamefont {Kousouris}}, \bibinfo
  {author} {\bibfnamefont {M.}~\bibnamefont {Kuster}}, \bibinfo {author}
  {\bibfnamefont {B.}~\bibnamefont {Laki\ifmmode~\acute{c}\else \'{c}\fi{}}},
  \bibinfo {author} {\bibfnamefont {C.}~\bibnamefont {Lasseur}}, \bibinfo
  {author} {\bibfnamefont {A.}~\bibnamefont {Liolios}}, \bibinfo {author}
  {\bibfnamefont {A.}~\bibnamefont {Ljubi\ifmmode \check{c}\else
  \v{c}\fi{}i\ifmmode~\acute{c}\else \'{c}\fi{}}}, \bibinfo {author}
  {\bibfnamefont {G.}~\bibnamefont {Lutz}}, \bibinfo {author} {\bibfnamefont
  {G.}~\bibnamefont {Luz\'on}}, \bibinfo {author} {\bibfnamefont {D.~W.}\
  \bibnamefont {Miller}}, \bibinfo {author} {\bibfnamefont {A.}~\bibnamefont
  {Morales}}, \bibinfo {author} {\bibfnamefont {J.}~\bibnamefont {Morales}},
  \bibinfo {author} {\bibfnamefont {M.}~\bibnamefont {Mutterer}}, \bibinfo
  {author} {\bibfnamefont {A.}~\bibnamefont {Nikolaidis}}, \bibinfo {author}
  {\bibfnamefont {A.}~\bibnamefont {Ortiz}}, \bibinfo {author} {\bibfnamefont
  {T.}~\bibnamefont {Papaevangelou}}, \bibinfo {author} {\bibfnamefont
  {A.}~\bibnamefont {Placci}}, \bibinfo {author} {\bibfnamefont
  {G.}~\bibnamefont {Raffelt}}, \bibinfo {author} {\bibfnamefont
  {J.}~\bibnamefont {Ruz}}, \bibinfo {author} {\bibfnamefont {H.}~\bibnamefont
  {Riege}}, \bibinfo {author} {\bibfnamefont {M.~L.}\ \bibnamefont {Sarsa}},
  \bibinfo {author} {\bibfnamefont {I.}~\bibnamefont {Savvidis}}, \bibinfo
  {author} {\bibfnamefont {W.}~\bibnamefont {Serber}}, \bibinfo {author}
  {\bibfnamefont {P.}~\bibnamefont {Serpico}}, \bibinfo {author} {\bibfnamefont
  {Y.}~\bibnamefont {Semertzidis}}, \bibinfo {author} {\bibfnamefont
  {L.}~\bibnamefont {Stewart}}, \bibinfo {author} {\bibfnamefont {J.~D.}\
  \bibnamefont {Vieira}}, \bibinfo {author} {\bibfnamefont {J.}~\bibnamefont
  {Villar}}, \bibinfo {author} {\bibfnamefont {L.}~\bibnamefont {Walckiers}},\
  and\ \bibinfo {author} {\bibfnamefont {K.}~\bibnamefont {Zachariadou}}
  (\bibinfo {collaboration} {CAST Collaboration}),\ }\bibfield  {title}
  {\bibinfo {title} {First results from the cern axion solar telescope},\
  }\href {https://doi.org/10.1103/PhysRevLett.94.121301} {\bibfield  {journal}
  {\bibinfo  {journal} {Phys. Rev. Lett.}\ }\textbf {\bibinfo {volume} {94}},\
  \bibinfo {pages} {121301} (\bibinfo {year} {2005})}\BibitemShut {NoStop}%
\bibitem [{\citenamefont {Collar}\ \emph {et~al.}(2003)\citenamefont {Collar}
  \emph {et~al.}}]{CAST:2003idc}%
  \BibitemOpen
  \bibfield  {author} {\bibinfo {author} {\bibfnamefont {J.~I.}\ \bibnamefont
  {Collar}} \emph {et~al.} (\bibinfo {collaboration} {CAST}),\ }\bibfield
  {title} {\bibinfo {title} {{CAST: A Search for solar axions at CERN}},\ }in\
  \href@noop {} {\emph {\bibinfo {booktitle} {{Conference on Astronomical
  Telescopes and Instrumenation}}}}\ (\bibinfo {year} {2003})\ \Eprint
  {https://arxiv.org/abs/hep-ex/0304024} {arXiv:hep-ex/0304024} \BibitemShut
  {NoStop}%
\bibitem [{\citenamefont {Armengaud}\ \emph {et~al.}(2019)\citenamefont
  {Armengaud} \emph {et~al.}}]{IAXO:2019mpb}%
  \BibitemOpen
  \bibfield  {author} {\bibinfo {author} {\bibfnamefont {E.}~\bibnamefont
  {Armengaud}} \emph {et~al.} (\bibinfo {collaboration} {IAXO}),\ }\bibfield
  {title} {\bibinfo {title} {{Physics potential of the International Axion
  Observatory (IAXO)}},\ }\href {https://doi.org/10.1088/1475-7516/2019/06/047}
  {\bibfield  {journal} {\bibinfo  {journal} {JCAP}\ }\textbf {\bibinfo
  {volume} {06}},\ \bibinfo {pages} {047}},\ \Eprint
  {https://arxiv.org/abs/1904.09155} {arXiv:1904.09155 [hep-ph]} \BibitemShut
  {NoStop}%
\bibitem [{\citenamefont {Coti~Zelati}\ \emph {et~al.}(2017)\citenamefont
  {Coti~Zelati}, \citenamefont {Rea}, \citenamefont {Pons}, \citenamefont
  {Campana},\ and\ \citenamefont {Esposito}}]{10.1093/mnras/stx2679}%
  \BibitemOpen
  \bibfield  {author} {\bibinfo {author} {\bibfnamefont {F.}~\bibnamefont
  {Coti~Zelati}}, \bibinfo {author} {\bibfnamefont {N.}~\bibnamefont {Rea}},
  \bibinfo {author} {\bibfnamefont {J.~A.}\ \bibnamefont {Pons}}, \bibinfo
  {author} {\bibfnamefont {S.}~\bibnamefont {Campana}},\ and\ \bibinfo {author}
  {\bibfnamefont {P.}~\bibnamefont {Esposito}},\ }\bibfield  {title} {\bibinfo
  {title} {{Systematic study of magnetar outbursts}},\ }\href
  {https://doi.org/10.1093/mnras/stx2679} {\bibfield  {journal} {\bibinfo
  {journal} {Monthly Notices of the Royal Astronomical Society}\ }\textbf
  {\bibinfo {volume} {474}},\ \bibinfo {pages} {961} (\bibinfo {year}
  {2017})},\ \Eprint
  {https://arxiv.org/abs/https://academic.oup.com/mnras/article-pdf/474/1/961/22370772/stx2679.pdf}
  {https://academic.oup.com/mnras/article-pdf/474/1/961/22370772/stx2679.pdf}
  \BibitemShut {NoStop}%
\bibitem [{\citenamefont {Armengaud}\ \emph {et~al.}(2018)\citenamefont
  {Armengaud} \emph {et~al.}}]{EDELWEISS:2018tde}%
  \BibitemOpen
  \bibfield  {author} {\bibinfo {author} {\bibfnamefont {E.}~\bibnamefont
  {Armengaud}} \emph {et~al.} (\bibinfo {collaboration} {EDELWEISS}),\
  }\bibfield  {title} {\bibinfo {title} {{Searches for electron interactions
  induced by new physics in the EDELWEISS-III Germanium bolometers}},\ }\href
  {https://doi.org/10.1103/PhysRevD.98.082004} {\bibfield  {journal} {\bibinfo
  {journal} {Phys. Rev. D}\ }\textbf {\bibinfo {volume} {98}},\ \bibinfo
  {pages} {082004} (\bibinfo {year} {2018})},\ \Eprint
  {https://arxiv.org/abs/1808.02340} {arXiv:1808.02340 [hep-ex]} \BibitemShut
  {NoStop}%
\bibitem [{\citenamefont {Akerib}\ \emph {et~al.}(2017)\citenamefont {Akerib}
  \emph {et~al.}}]{LUX:2017glr}%
  \BibitemOpen
  \bibfield  {author} {\bibinfo {author} {\bibfnamefont {D.~S.}\ \bibnamefont
  {Akerib}} \emph {et~al.} (\bibinfo {collaboration} {LUX}),\ }\bibfield
  {title} {\bibinfo {title} {{First Searches for Axions and Axionlike Particles
  with the LUX Experiment}},\ }\href
  {https://doi.org/10.1103/PhysRevLett.118.261301} {\bibfield  {journal}
  {\bibinfo  {journal} {Phys. Rev. Lett.}\ }\textbf {\bibinfo {volume} {118}},\
  \bibinfo {pages} {261301} (\bibinfo {year} {2017})},\ \Eprint
  {https://arxiv.org/abs/1704.02297} {arXiv:1704.02297 [astro-ph.CO]}
  \BibitemShut {NoStop}%
\bibitem [{\citenamefont {Fu}\ \emph {et~al.}(2017)\citenamefont {Fu} \emph
  {et~al.}}]{PandaX:2017ock}%
  \BibitemOpen
  \bibfield  {author} {\bibinfo {author} {\bibfnamefont {C.}~\bibnamefont {Fu}}
  \emph {et~al.} (\bibinfo {collaboration} {PandaX}),\ }\bibfield  {title}
  {\bibinfo {title} {{Limits on Axion Couplings from the First 80 Days of Data
  of the PandaX-II Experiment}},\ }\href
  {https://doi.org/10.1103/PhysRevLett.119.181806} {\bibfield  {journal}
  {\bibinfo  {journal} {Phys. Rev. Lett.}\ }\textbf {\bibinfo {volume} {119}},\
  \bibinfo {pages} {181806} (\bibinfo {year} {2017})},\ \Eprint
  {https://arxiv.org/abs/1707.07921} {arXiv:1707.07921 [hep-ex]} \BibitemShut
  {NoStop}%
\bibitem [{\citenamefont {Gondolo}\ and\ \citenamefont
  {Raffelt}(2009)}]{Gondolo:2008dd}%
  \BibitemOpen
  \bibfield  {author} {\bibinfo {author} {\bibfnamefont {P.}~\bibnamefont
  {Gondolo}}\ and\ \bibinfo {author} {\bibfnamefont {G.~G.}\ \bibnamefont
  {Raffelt}},\ }\bibfield  {title} {\bibinfo {title} {{Solar neutrino limit on
  axions and keV-mass bosons}},\ }\href
  {https://doi.org/10.1103/PhysRevD.79.107301} {\bibfield  {journal} {\bibinfo
  {journal} {Phys. Rev. D}\ }\textbf {\bibinfo {volume} {79}},\ \bibinfo
  {pages} {107301} (\bibinfo {year} {2009})},\ \Eprint
  {https://arxiv.org/abs/0807.2926} {arXiv:0807.2926 [astro-ph]} \BibitemShut
  {NoStop}%
\bibitem [{\citenamefont {Capozzi}\ and\ \citenamefont
  {Raffelt}(2020)}]{Capozzi:2020cbu}%
  \BibitemOpen
  \bibfield  {author} {\bibinfo {author} {\bibfnamefont {F.}~\bibnamefont
  {Capozzi}}\ and\ \bibinfo {author} {\bibfnamefont {G.}~\bibnamefont
  {Raffelt}},\ }\bibfield  {title} {\bibinfo {title} {{Axion and neutrino
  bounds improved with new calibrations of the tip of the red-giant branch
  using geometric distance determinations}},\ }\href
  {https://doi.org/10.1103/PhysRevD.102.083007} {\bibfield  {journal} {\bibinfo
   {journal} {Phys. Rev. D}\ }\textbf {\bibinfo {volume} {102}},\ \bibinfo
  {pages} {083007} (\bibinfo {year} {2020})},\ \Eprint
  {https://arxiv.org/abs/2007.03694} {arXiv:2007.03694 [astro-ph.SR]}
  \BibitemShut {NoStop}%
\bibitem [{\citenamefont {Straniero}\ \emph {et~al.}(2020)\citenamefont
  {Straniero}, \citenamefont {Pallanca}, \citenamefont {Dalessandro},
  \citenamefont {Dominguez}, \citenamefont {Ferraro}, \citenamefont
  {Giannotti}, \citenamefont {Mirizzi},\ and\ \citenamefont
  {Piersanti}}]{Straniero:2020iyi}%
  \BibitemOpen
  \bibfield  {author} {\bibinfo {author} {\bibfnamefont {O.}~\bibnamefont
  {Straniero}}, \bibinfo {author} {\bibfnamefont {C.}~\bibnamefont {Pallanca}},
  \bibinfo {author} {\bibfnamefont {E.}~\bibnamefont {Dalessandro}}, \bibinfo
  {author} {\bibfnamefont {I.}~\bibnamefont {Dominguez}}, \bibinfo {author}
  {\bibfnamefont {F.~R.}\ \bibnamefont {Ferraro}}, \bibinfo {author}
  {\bibfnamefont {M.}~\bibnamefont {Giannotti}}, \bibinfo {author}
  {\bibfnamefont {A.}~\bibnamefont {Mirizzi}},\ and\ \bibinfo {author}
  {\bibfnamefont {L.}~\bibnamefont {Piersanti}},\ }\bibfield  {title} {\bibinfo
  {title} {{The RGB tip of galactic globular clusters and the revision of the
  axion-electron coupling bound}},\ }\href
  {https://doi.org/10.1051/0004-6361/202038775} {\bibfield  {journal} {\bibinfo
   {journal} {Astron. Astrophys.}\ }\textbf {\bibinfo {volume} {644}},\
  \bibinfo {pages} {A166} (\bibinfo {year} {2020})},\ \Eprint
  {https://arxiv.org/abs/2010.03833} {arXiv:2010.03833 [astro-ph.SR]}
  \BibitemShut {NoStop}%
\bibitem [{\citenamefont {O'Hare}(2020)}]{ciaran_o_hare_2020_3932430}%
  \BibitemOpen
  \bibfield  {author} {\bibinfo {author} {\bibfnamefont {C.}~\bibnamefont
  {O'Hare}},\ }\href {https://doi.org/10.5281/zenodo.3932430} {\bibinfo {title}
  {cajohare/axionlimits: Axionlimits}} (\bibinfo {year} {2020})\BibitemShut
  {NoStop}%
\bibitem [{\citenamefont {Luzio}\ \emph {et~al.}(2021)\citenamefont {Luzio},
  \citenamefont {Fedele}, \citenamefont {Giannotti}, \citenamefont {Mescia},\
  and\ \citenamefont {Nardi}}]{diluzio2021stellar}%
  \BibitemOpen
  \bibfield  {author} {\bibinfo {author} {\bibfnamefont {L.~D.}\ \bibnamefont
  {Luzio}}, \bibinfo {author} {\bibfnamefont {M.}~\bibnamefont {Fedele}},
  \bibinfo {author} {\bibfnamefont {M.}~\bibnamefont {Giannotti}}, \bibinfo
  {author} {\bibfnamefont {F.}~\bibnamefont {Mescia}},\ and\ \bibinfo {author}
  {\bibfnamefont {E.}~\bibnamefont {Nardi}},\ }\href@noop {} {\bibinfo {title}
  {Stellar evolution confronts axion models}} (\bibinfo {year} {2021}),\
  \Eprint {https://arxiv.org/abs/2109.10368} {arXiv:2109.10368 [hep-ph]}
  \BibitemShut {NoStop}%
\bibitem [{\citenamefont {Veigele}(1973)}]{VEIGELE197351}%
  \BibitemOpen
  \bibfield  {author} {\bibinfo {author} {\bibfnamefont {W.}~\bibnamefont
  {Veigele}},\ }\bibfield  {title} {\bibinfo {title} {Photon cross sections
  from 0.1 kev to 1 mev for elements z = 1 to z = 94},\ }\href
  {https://doi.org/https://doi.org/10.1016/S0092-640X(73)80015-4} {\bibfield
  {journal} {\bibinfo  {journal} {Atomic Data and Nuclear Data Tables}\
  }\textbf {\bibinfo {volume} {5}},\ \bibinfo {pages} {51} (\bibinfo {year}
  {1973})}\BibitemShut {NoStop}%
\bibitem [{\citenamefont {Arisaka}\ \emph {et~al.}(2013)\citenamefont
  {Arisaka}, \citenamefont {Beltrame}, \citenamefont {Ghag}, \citenamefont
  {Kaidi}, \citenamefont {Lung}, \citenamefont {Lyashenko}, \citenamefont
  {Peccei}, \citenamefont {Smith},\ and\ \citenamefont {Ye}}]{Arisaka:2012pb}%
  \BibitemOpen
  \bibfield  {author} {\bibinfo {author} {\bibfnamefont {K.}~\bibnamefont
  {Arisaka}}, \bibinfo {author} {\bibfnamefont {P.}~\bibnamefont {Beltrame}},
  \bibinfo {author} {\bibfnamefont {C.}~\bibnamefont {Ghag}}, \bibinfo {author}
  {\bibfnamefont {J.}~\bibnamefont {Kaidi}}, \bibinfo {author} {\bibfnamefont
  {K.}~\bibnamefont {Lung}}, \bibinfo {author} {\bibfnamefont {A.}~\bibnamefont
  {Lyashenko}}, \bibinfo {author} {\bibfnamefont {R.~D.}\ \bibnamefont
  {Peccei}}, \bibinfo {author} {\bibfnamefont {P.}~\bibnamefont {Smith}},\ and\
  \bibinfo {author} {\bibfnamefont {K.}~\bibnamefont {Ye}},\ }\bibfield
  {title} {\bibinfo {title} {{Expected Sensitivity to Galactic/Solar Axions and
  Bosonic Super-WIMPs based on the Axio-electric Effect in Liquid Xenon Dark
  Matter Detectors}},\ }\href
  {https://doi.org/10.1016/j.astropartphys.2012.12.009} {\bibfield  {journal}
  {\bibinfo  {journal} {Astropart. Phys.}\ }\textbf {\bibinfo {volume} {44}},\
  \bibinfo {pages} {59} (\bibinfo {year} {2013})},\ \Eprint
  {https://arxiv.org/abs/1209.3810} {arXiv:1209.3810 [astro-ph.CO]}
  \BibitemShut {NoStop}%
\bibitem [{\citenamefont {Arisaka}\ \emph {et~al.}(2009)\citenamefont
  {Arisaka}, \citenamefont {Wang}, \citenamefont {Smith}, \citenamefont
  {Cline}, \citenamefont {Teymourian}, \citenamefont {Brown}, \citenamefont
  {Ooi}, \citenamefont {Aharoni}, \citenamefont {Lam}, \citenamefont {Lung},
  \citenamefont {Davies},\ and\ \citenamefont {Price}}]{ARISAKA200963}%
  \BibitemOpen
  \bibfield  {author} {\bibinfo {author} {\bibfnamefont {K.}~\bibnamefont
  {Arisaka}}, \bibinfo {author} {\bibfnamefont {H.}~\bibnamefont {Wang}},
  \bibinfo {author} {\bibfnamefont {P.}~\bibnamefont {Smith}}, \bibinfo
  {author} {\bibfnamefont {D.}~\bibnamefont {Cline}}, \bibinfo {author}
  {\bibfnamefont {A.}~\bibnamefont {Teymourian}}, \bibinfo {author}
  {\bibfnamefont {E.}~\bibnamefont {Brown}}, \bibinfo {author} {\bibfnamefont
  {W.}~\bibnamefont {Ooi}}, \bibinfo {author} {\bibfnamefont {D.}~\bibnamefont
  {Aharoni}}, \bibinfo {author} {\bibfnamefont {C.}~\bibnamefont {Lam}},
  \bibinfo {author} {\bibfnamefont {K.}~\bibnamefont {Lung}}, \bibinfo {author}
  {\bibfnamefont {S.}~\bibnamefont {Davies}},\ and\ \bibinfo {author}
  {\bibfnamefont {M.}~\bibnamefont {Price}},\ }\bibfield  {title} {\bibinfo
  {title} {Xax: A multi-ton, multi-target detection system for dark matter,
  double beta decay and pp solar neutrinos},\ }\href
  {https://doi.org/https://doi.org/10.1016/j.astropartphys.2008.11.007}
  {\bibfield  {journal} {\bibinfo  {journal} {Astroparticle Physics}\ }\textbf
  {\bibinfo {volume} {31}},\ \bibinfo {pages} {63} (\bibinfo {year}
  {2009})}\BibitemShut {NoStop}%
\bibitem [{\citenamefont {Dong}\ \emph {et~al.}(2016)\citenamefont {Dong} \emph
  {et~al.}}]{Dong:2015oea}%
  \BibitemOpen
  \bibfield  {author} {\bibinfo {author} {\bibfnamefont {S.}~\bibnamefont
  {Dong}} \emph {et~al.},\ }\bibfield  {title} {\bibinfo {title} {{ASASSN-15lh:
  A Highly Super-Luminous Supernova}},\ }\href
  {https://doi.org/10.1126/science.aac9613} {\bibfield  {journal} {\bibinfo
  {journal} {Science}\ }\textbf {\bibinfo {volume} {351}},\ \bibinfo {pages}
  {257} (\bibinfo {year} {2016})},\ \Eprint {https://arxiv.org/abs/1507.03010}
  {arXiv:1507.03010 [astro-ph.HE]} \BibitemShut {NoStop}%
\bibitem [{\citenamefont {Kankare}\ \emph {et~al.}(2017)\citenamefont {Kankare}
  \emph {et~al.}}]{Kankare:2017rtx}%
  \BibitemOpen
  \bibfield  {author} {\bibinfo {author} {\bibfnamefont {E.}~\bibnamefont
  {Kankare}} \emph {et~al.},\ }\bibfield  {title} {\bibinfo {title} {{A
  population of highly energetic transient events in the centres of active
  galaxies}},\ }\href {https://doi.org/10.1038/s41550-017-0290-2} {\bibfield
  {journal} {\bibinfo  {journal} {Nature Astron.}\ }\textbf {\bibinfo {volume}
  {1}},\ \bibinfo {pages} {865} (\bibinfo {year} {2017})},\ \Eprint
  {https://arxiv.org/abs/1711.04577} {arXiv:1711.04577 [astro-ph.HE]}
  \BibitemShut {NoStop}%
\bibitem [{\citenamefont {Denissenya}\ \emph {et~al.}(2021)\citenamefont
  {Denissenya}, \citenamefont {Grossan},\ and\ \citenamefont
  {Linder}}]{PhysRevD.104.023007}%
  \BibitemOpen
  \bibfield  {author} {\bibinfo {author} {\bibfnamefont {M.}~\bibnamefont
  {Denissenya}}, \bibinfo {author} {\bibfnamefont {B.}~\bibnamefont
  {Grossan}},\ and\ \bibinfo {author} {\bibfnamefont {E.~V.}\ \bibnamefont
  {Linder}},\ }\bibfield  {title} {\bibinfo {title} {Distinguishing time
  clustering of astrophysical bursts},\ }\href
  {https://doi.org/10.1103/PhysRevD.104.023007} {\bibfield  {journal} {\bibinfo
   {journal} {Phys. Rev. D}\ }\textbf {\bibinfo {volume} {104}},\ \bibinfo
  {pages} {023007} (\bibinfo {year} {2021})}\BibitemShut {NoStop}%
\bibitem [{\citenamefont {Aprile}\ \emph {et~al.}(2020)\citenamefont {Aprile}
  \emph {et~al.}}]{XENON:2020rca}%
  \BibitemOpen
  \bibfield  {author} {\bibinfo {author} {\bibfnamefont {E.}~\bibnamefont
  {Aprile}} \emph {et~al.} (\bibinfo {collaboration} {XENON}),\ }\bibfield
  {title} {\bibinfo {title} {{Excess electronic recoil events in XENON1T}},\
  }\href {https://doi.org/10.1103/PhysRevD.102.072004} {\bibfield  {journal}
  {\bibinfo  {journal} {Phys. Rev. D}\ }\textbf {\bibinfo {volume} {102}},\
  \bibinfo {pages} {072004} (\bibinfo {year} {2020})},\ \Eprint
  {https://arxiv.org/abs/2006.09721} {arXiv:2006.09721 [hep-ex]} \BibitemShut
  {NoStop}%
\bibitem [{\citenamefont {Di~Luzio}\ \emph {et~al.}(2020)\citenamefont
  {Di~Luzio}, \citenamefont {Fedele}, \citenamefont {Giannotti}, \citenamefont
  {Mescia},\ and\ \citenamefont {Nardi}}]{DiLuzio:2020jjp}%
  \BibitemOpen
  \bibfield  {author} {\bibinfo {author} {\bibfnamefont {L.}~\bibnamefont
  {Di~Luzio}}, \bibinfo {author} {\bibfnamefont {M.}~\bibnamefont {Fedele}},
  \bibinfo {author} {\bibfnamefont {M.}~\bibnamefont {Giannotti}}, \bibinfo
  {author} {\bibfnamefont {F.}~\bibnamefont {Mescia}},\ and\ \bibinfo {author}
  {\bibfnamefont {E.}~\bibnamefont {Nardi}},\ }\bibfield  {title} {\bibinfo
  {title} {{Solar axions cannot explain the XENON1T excess}},\ }\href
  {https://doi.org/10.1103/PhysRevLett.125.131804} {\bibfield  {journal}
  {\bibinfo  {journal} {Phys. Rev. Lett.}\ }\textbf {\bibinfo {volume} {125}},\
  \bibinfo {pages} {131804} (\bibinfo {year} {2020})},\ \Eprint
  {https://arxiv.org/abs/2006.12487} {arXiv:2006.12487 [hep-ph]} \BibitemShut
  {NoStop}%
\end{thebibliography}%

\end{document}